\documentclass[
reprint,
 superscriptaddress,
 amsmath,amssymb,
prb,
floatfix,
]{revtex4-2}
\usepackage{amsmath}
\usepackage{graphicx}% Include figure files
\usepackage{dcolumn}% Align table columns on decimal point
\usepackage{comment}
\usepackage{bm}
\usepackage{makecell}
\usepackage{tikz}
\usepackage{color}
\usepackage{braket}
\usepackage{bbold}
\usepackage{cleveref}
\usepackage{tabularx}
\usepackage{booktabs}
\usepackage[T1]{fontenc}
\usepackage[utf8]{inputenc}
\usepackage{caption}
\usepackage{cellspace}

\begin{document}

\title{Non-adiabatic Ehrenfest dynamics with norm-conserving and ultra-soft pseudo-potentials with nuclear velocity corrections on the atomic orbitals within the Projector Augmented Wave Method framework}

\author{Paolo Fachin}
\affiliation{Dipartimento di Fisica, Sapienza - Università di Roma, Roma, Italy}%
\author{Francesco Macheda}
\affiliation{Dipartimento di Fisica, Sapienza - Università di Roma, Roma, Italy}%
\affiliation{Dipartimento di Scienze e Metodi dell’Ingegneria, Università di Modena e Reggio Emilia, Reggio Emilia, Italy}
\author{Paolo Barone}%
\affiliation{Dipartimento di Fisica, Sapienza - Università di Roma, Roma, Italy}
\affiliation{CNR-SPIN, Area della Ricerca di Tor Vergata, Via del Fosso del Cavaliere 100, I-00133 Rome, Italy}
\author{Francesco Mauri}
\affiliation{Dipartimento di Fisica, Sapienza - Università di Roma, Roma, Italy}%

\newcommand{\citeSupp}[0]{Note1}

\begin{abstract}
We derive the first-principles Ehrenfest molecular dynamics describing non-adiabatic processes with the inclusion of the nuclear-velocity-dependent phases (also known as electron-translation factors) on the atomic-orbital basis. These phases, appearing when nuclei are treated dynamically, affect effective Hamiltonians constructed from localised orbitals. In this work,  we focus on the effects in the first-principles pseudo-potential Hamiltonian, both for the norm-conserving and ultra-soft cases, derived within the Projector-Augmented-Wave (PAW) method framework. Peierls-like phases depending on the nuclear velocities appear in the non-local part of the potential, while additional nuclear velocity and acceleration-dependent corrections appear in the ultra-soft pseudo-potential case. The use of velocity-including atomic orbital basis enables a Galilean-invariant description of the non-adiabatic Ehrenfest molecular dynamics, removing spurious non-adiabatic couplings that arise from neglecting the nuclear velocity phases in the atomic orbitals. 
\end{abstract}

\maketitle

\section{Introduction}
Effective electronic Hamiltonians are widely used in condensed matter physics to describe the low-energy spectrum of the system and the related physical properties \cite{Grosso2013}. Within the \textit{ab-initio} framework, the use of pseudo-potentials enables the achievement of large computational gains with respect to all-electron calculations \cite{PhysRevLett.43.1494,PhysRevB.26.4199,PhysRevB.43.1993,PhysRevB.41.7892,PhysRevB.50.17953,PhysRevB.88.085117,PhysRevLett.48.1425, VANSETTEN201839}. In the Born-Oppenheimer approximation, effective electronic Hamiltonians are usually derived starting from the all-electron one, at fixed  nuclei, and excluding non-adiabatic effects. Even when non-adiabatic effects are included in the nuclear dynamics, the effective electronic Hamiltonians are still usually constructed using atomic-like orbitals as if the nuclei were at rest, or at most considering the electronic wavefunction rigidly shifting with the nuclei. This approach introduces discrepancies between the dynamics of the effective models and of the all-electron Hamiltonian. For example, in the context of the non-adiabatic molecular dynamics, spurious couplings arise in the effective Hamiltonian governing the temporal evolution of the electronic systems \cite{10.1063/1.3665031, doi:10.1021/jz3006173, doi:10.1021/jp9906839, PhysRevA.62.022703}. They manifest in the paradox that a motion at constant velocity of an atom can induce electronic transitions \cite{10.1063/1.3665031, doi:10.1021/jz3006173}, clearly breaking the principle of Galilean invariance. Analogously, the breaking of Galilean invariance manifests in the violation of nonadiabatic sum rules relating  Born effective charges and the interatomic force constant matrix to the frequency-dependent electromagnetic susceptibility, when calculated with pseudo-potential Hamiltonians obtained rigidly shifting the electronic orbitals with the nuclei, as opposed to all-electron ones, as recently pointed out in Ref. \cite{PhysRevLett.136.196401}.  

These issues arise from neglecting the nuclear-velocity dependent phases in the atomic-like orbitals \cite{ 10.1063/1.3665031,doi:10.1021/jz3006173, z497-65ks, PhysRevLett.136.196401, doi:10.1021/jz3006173, 10.1063/1.3665031, 10.1063/5.0160965}, often called "electron-translation factors (ETFs)" in the literature. The use of nuclear-velocity-including atomic orbital basis has a long history in the study of atomic collisions, where it was first introduced \cite{Bates1958_electroncapture} and saw several developments \cite{LFErrea_1994, PhysRevA.50.418, PhysRevA.35.70, PhysRevA.85.012702, RJAllan_1985, RJAllan1985bis,PhysRevA.18.117, PhysRevA.82.060701, z497-65ks, PhysRevA.82.060701, PhysRevA.23.2301, PhysRevLett.80.3029}. In the study of the properties of molecules and solids, velocity-including atomic orbitals were first introduced by Nafie to assess vibrational circular dichroism (VCD) in Ref. \cite{10.1063/1.462668} using the complete adiabatic approach \cite{10.1063/1.445588}.  Nuclear velocity-dependent phases enable non-zero electronic currents and magnetic dipole moments, which, conversely, vanish for the standard real-valued Born-Oppenheimer electronic wavefunctions, resulting in a vanishing electronic contribution to VCD \cite{10.1063/1.4928578, 10.1063/1.4747540}.  In this framework, nuclear velocity perturbation theory (NVPT) has been developed and implemented, mainly for the purpose of determining VCD spectra \cite{2023_Ditler_Mattiat_Luber, doi:10.1021/ct400700c, doi:10.1021/acs.jctc.2c00006, doi:10.1021/acs.jpca.5c01344}.  It is worth noticing that the exact factorisation method \cite{PhysRevLett.105.123002}  encodes these effects, as it can be appreciated in the classical limit for the nuclear wavefunction \cite{10.1063/1.4928578}. 

Besides being essential for correctly describing VCD spectra, the use of nuclear-velocity-including atomic orbital basis allows for the removal of spurious electronic transitions in the dynamics. This has been shown, at the linear order in the nuclear velocity, in the context of a time-dependent Hartree-Fock dynamics in Refs. \cite{doi:10.1021/jp9906839, PhysRevA.62.022703, REYES2002441, doi:10.1021/jp505767b} and for localised atomic orbitals in Refs. \cite{10.1063/1.3665031, doi:10.1021/jz3006173, 10.1063/5.0160965}, as well as in the context of TDDFT linear response theory in Refs. \cite{doi:10.1021/acs.jctc.5c01960,10.1063/1.4906941}. 
Furthermore, they restore the frequency-dependent sum rules in the framework of pseudopotential approximation, correctly accounting for nonadiabatic couplings and frequency-dependent vibrational responses within linear-response theory \cite{PhysRevLett.136.196401}. Nonetheless, nuclear velocity effects on the wavefunctions are still neglected in current implementations of first-principles non-adiabatic molecular dynamics \cite{10.1063/1.3700800, PhysRevB.73.035408, 10.1063/5.0182685, doi:10.1021/ct400641n, doi:10.1021/acs.jpclett.0c03853, doi:10.1021/acs.jpclett.0c03080,doi:10.1021/acs.jctc.5c01082, 10.1063/5.0230570}. 

In this work, we determine the first-principles non-adiabatic semiclassical Ehrenfest dynamics - where nuclei are classical and electrons are quantum mechanical particles \cite{Todorov_2001} -  within the Projector Augmented Wave (PAW) method framework for both the ultra-soft and norm-conserving cases, using nuclear-velocity-including atomic orbitals.  %Generalisation to quantum nuclei can be obtained  as this work.  
Our approach enables the removal of the spurious contribution in the non-adiabatic Ehrenfest dynamics, recovering the physical results of the all-electron dynamics with the pseudopotential one, where full Galilean invariance is respected. These nuclear velocity dependences manifest as Peierls-like phases appearing in the non-local part of the pseudopotential, capturing the response at any order in the nuclear velocity, as already discussed in Ref. \cite{PhysRevLett.136.196401}. We generalise the results of Ref. \cite{PhysRevLett.136.196401} to the case of ultrasoft pseudo-potentials and we consider the contributions from the temporal derivatives of the PAW transformation. They originate additional nuclear velocity and acceleration dependent terms, relevant in the ultrasoft case, always neglected in the previous literature.

The paper is organised as follows: in Sec. \ref{sec:I} we introduce the velocity-including atomic orbitals, studying the electronic problem for an isolated nucleus in motion, while in Sec. \ref{sec:II} we present the semiclassical Ehrenfest Lagrangian formalism used to derive the effective Hamiltonians and the equations of motion for velocity-including atomic orbitals. The standard PAW method is explained in Section \ref{sec:PAW}, showing the adiabatic Hamiltonian and the related responses. These instruments allow us to review the currently implemented non-adiabatic PAW Ehrenfest dynamics \cite{PhysRevB.73.035408, 10.1063/1.3700800, 10.1063/5.0182685} in \ref{subsec:dynamics_PAW_standard}. Conversely, Section \ref{sec:PAW_vel} is devoted to explaining the PAW method with the velocity-including atomic orbitals, which enables the proper description of the non-adiabatic Ehrenfest dynamics \ref{sec:Ehrnefest_PAW_vel}. Finally,
we draw our conclusions in Section \ref{sec:conclusion}. The Hamiltonians, the electronic temporal dynamics and the conserved energies are compared with the literature in Table \ref{tab:hamiltonians_dynamics}, that summarises the main findings of this work. 

\section{Velocity-including atomic orbitals}
\label{sec:I}
 Localised atomic orbital bases are usually constructed from the orbitals of isolated atoms with the nucleus fixed in its position. In this paragraph, we establish a localised atomic orbital basis for the case of moving nuclei. 
 
Consider an isolated atom $s$ with the nucleus at rest, which, without loss of generality, is located at the origin. The electrons are described by a single particle all-electron Hamiltonian 
\begin{equation}
\hat{H}_{\mathbf{0}}^{\mathrm{AE}}(\hat{\mathbf{r}},\hat{\mathbf{p}})=\frac{\hat{\mathbf{p}}^2}{2m}+V_s(\hat{\mathbf{r}}).
\label{eq:all-electron_hamiltonian_static}
\end{equation}
where the hat $\hat{}$ denotes operators, $V_s(\hat{\mathbf{r}})$ is the effective potential acting on the electron, the subscript $s$ indicates the dependence of the potential on the atomic properties of $s$. The Hamiltonian is diagonalised by the atomic orbitals $\ket{\phi^{\rm \mathbf{0}}_{si}}$ with energy $E_{si}$, where $i$ indicates the electronic quantum number and the superscript $\mathbf{0}$ indicates that the orbitals $\ket{\phi^{\rm \mathbf{0}}_{si}}$ are solutions for the atom at rest at the origin,
\begin{equation}
    \hat{H}_{\mathbf{0}}^{\mathrm{AE}}(\hat{\mathbf{r}},\hat{\mathbf{p}})\ket{\phi^{\rm \mathbf{0}}_{si}}=E_{si}\ket{\phi^{\rm \mathbf{0}}_{si}}.
\end{equation}
Usually, localised atomic orbital bases are constructed using $\ket{\phi^{\rm \mathbf{0}}_{si}}$ orbitals, translated to the positions of atoms of the system
\begin{equation}
    \ket{\phi^{\mathbf{R}_{s}}_{si}}=\hat{T}_{\mathbf{R}_{s}}\ket{\phi^{\mathbf{0}}_{si}},
\end{equation}
that form the basis set $\{\ket{\phi^{\mathbf{R}_{s}}_{si}}\}_{si}$. We remark that the translation operator  $\hat{T}_{\mathbf{R}_{s}}$ acts on the position eigenstates as $\hat{T}_{\mathbf{R}_s}\ket{\mathbf{r}}=\ket{\mathbf{r}+\mathbf{R}_s}$. The rigid translation of the orbitals assumes that nuclei are static. If the nuclei are moving, with their coordinate explicitly depending on time $\mathbf{R}_{s}(t)$, the rigidly translated atomic orbitals, in spite of their common use, should be replaced by atomic orbitals accounting for the nuclear motion. In the following, we derive this basis set.

Suppose for $t<0$ the $s$ nucleus, identified by the coordinate $\mathbf{R}_s(t)$, is at rest ($\mathbf{R}_s(t<0)=\bm{0}$) and that, at $t=0$, the nucleus starts moving with velocity $\dot{\mathbf{R}}_s(t)$ and acceleration $\ddot{\mathbf{R}}_s(t)$. At $t=0$ $\ket{\phi_{si}(t)}=\ket{\phi^{\rm \mathbf{0}}_{si}}$, then the time evolution of the atomic orbital $\ket{\phi_{si}(t)}$ for $t>0$ is determined by the Schrödinger equation with the Hamiltonian $\hat{H}^{\mathrm{AE}}(\hat{\mathbf{r}},\hat{\mathbf{p}};\mathbf{R}_s(t))$, that depends on time through the nuclear position $\mathbf{R}_s(t)$,
\begin{align}
    &i\hbar \frac{d\ket{\phi_{si}(t)}}{dt}=\hat{H}^{\mathrm{AE}}(\hat{\mathbf{r}},\hat{\mathbf{p}};\mathbf{R}_s(t)) \ket{\phi_{si}(t)},\label{eq:Schrodinger_moving_ion}\\
  &  \hat{H}^{\mathrm{AE}}(\hat{\mathbf{r}},\hat{\mathbf{p}};\mathbf{R}_s(t))=\frac{\hat{\mathbf{p}}^2}{2m}+V_s(\hat{\mathbf{r}}-\mathbf{R}_s(t)).
    \label{eq:all-electron_nuclear_hamiltonian_moving}
\end{align}
For $t>0$, we suppose that the solution of the above Schrödinger equation (Eq. \eqref{eq:Schrodinger_moving_ion}) is expressed in the form of
\begin{align}
    \ket{\phi_{si}(t)}&=e^{i\alpha_{s}(\hat{\mathbf{r}})}\hat{T}_{\mathbf{R}_{s}(t)}e^{i\Theta(t)}\ket{\phi'_{si} (t)},\label{eq:atomic_orbital_I}\\
    \alpha_{s}(\hat{\mathbf{r}})&=\frac{m}{\hbar}\dot{\mathbf{R}}_s(t)\cdot(\hat{\mathbf{r}}-\mathbf{R}_s(t)), \\
    \Theta(t)&=-\frac{1}{\hbar}\left(E_{si}t-\int_0^t dt' \frac{m|\dot{\mathbf{R}}_s(t')|^2}{2}\right). \nonumber
\end{align} 
We remark that the velocity-dependent phase $\alpha_{s}(\hat{\mathbf{r}})$ does not depend on the choice of the origin of the reference frame. Its time dependence is instantaneous. Conversely, the purely time-dependent phase $\Theta(t)$ and $\ket{\phi'_{si} (t)}$ depend on the system's history. In particular, the phase  $\Theta(t)$ depends on the temporal integration of the electron's kinetic energy moving at the velocity of the nucleus.  

To obtain the differential equation for $\ket{\phi'_{si} (t)}$, we plug Eq. \eqref{eq:atomic_orbital_I} into Eq. \eqref{eq:Schrodinger_moving_ion}.  The temporal derivative on the left-hand side is 
\begin{equation}
\begin{aligned}
  &\frac{d \ket{\phi_{si}(t)}}{dt}
    = e^{i\alpha_{s}(\hat{\mathbf{r}})}\hat{T}_{\mathbf{R}_{s}(t)}e^{i\Theta(t)}\Bigg[\frac{d \ket{\phi'_{si}(t)}}{dt}-\frac{i}{\hbar}\Bigg(E_{si}\\&+\dot{\mathbf{R}}_s(t)\cdot \hat{\mathbf{p}} -m \ddot{\mathbf{R}}_s(t)\cdot \hat{\mathbf{r}}
   +\frac{m|\dot{\mathbf{R}}_s(t)|^2}{2}\Bigg)\ket{\phi'_{si}(t)}\Bigg]
    \label{eq:derivative_velocity_including}
\end{aligned}
\end{equation}
where we used that $\hat{T}^{\dagger}_{\mathbf{R}_{s}(t)} \hat{\mathbf{r}}\hat{T}_{\mathbf{R}_{s}(t)}=\hat{\mathbf{r}}+\mathbf{R}_s(t)$.
On the right-hand side of Eq. \eqref{eq:Schrodinger_moving_ion}, the application of the Hamiltonian to $\ket{\phi_{si}(t)}$ corresponds to
 \begin{equation}
      \begin{aligned}
 & \hat{H}^{\mathrm{AE}}(\hat{\mathbf{r}},\hat{\mathbf{p}};\mathbf{R}_s(t)) e^{i\alpha_{s}(\hat{\mathbf{r}})}\hat{T}_{\mathbf{R}_{s}(t)}
   \\
   =&e^{i\alpha_{s}(\hat{\mathbf{r}})}
    \hat{T}_{\mathbf{R}_{s}(t)}\hat{H}^{\mathrm{AE}}(\hat{\mathbf{r}}+\mathbf{R}_s(t),\hat{\mathbf{p}}+m\dot{\mathbf{R}}_s(t);\mathbf{R}_s(t)) . \label{eq:Hamiltonian_isolated_atom_translated}
\end{aligned}
 \end{equation}
where we used $e^{-i\alpha_{s}(\hat{\mathbf{r}})}\hat{\mathbf{p}}e^{i\alpha_{s}(\hat{\mathbf{r}})}=\hat{\mathbf{p}}+m\dot{\mathbf{R}}_s(t)$. Conversely, the purely time-dependent phase factor commutes with the Hamiltonian. By putting together Eq. \eqref{eq:derivative_velocity_including} and Eq. \eqref{eq:Hamiltonian_isolated_atom_translated},  we obtain that the Schrödinger equation for moving nuclei of Eq. \eqref{eq:Schrodinger_moving_ion} is equivalent to the following equations for $\ket{\phi'_{si}(t)}$ with the boundary condition $\ket{\phi'_{si}(t=0)}=\ket{\phi^{\rm \mathbf{0}}_{si}}$
\begin{align}
    &i\hbar \frac{d\ket{\phi'_{si}(t)}}{dt}=\left(\hat{H}'^{\mathrm{AE}}(\hat{\mathbf{r}}, \hat{\mathbf{p}};\ddot{\mathbf{R}}_s(t))-E_{si}\right) \ket{\phi'_{si}(t)},\label{eq:Schrodinger_phi0_moving_ions}\\
  &\hat{H}'^{\mathrm{AE}}(\hat{\mathbf{r}}, \hat{\mathbf{p}};\ddot{\mathbf{R}}_s(t))=\frac{\hat{\mathbf{p}}^2}{2m}+V_s(\hat{\mathbf{r}})+m\ddot{\mathbf{R}}_s(t)\cdot \hat{\mathbf{r}}\label{eq:Hamiltonian_non_intertial_frame}.
\end{align}
$\hat{H}'^{\mathrm{AE}}(\hat{\mathbf{r}}, \hat{\mathbf{p}};\ddot{\mathbf{R}}_s(t))$ differs from the $\hat{H}_{\mathbf{0}}^{\mathrm{AE}}(\hat{\mathbf{r}}, \hat{\mathbf{p}})$ only by the term $m\ddot{\mathbf{R}}_s(t)\cdot \hat{\mathbf{r}}$, that can be interpreted as the non-inertial force acting on the electrons in the frame where the nucleus is at rest. Its effect on the Hamiltonian is analogous to a time-dependent electric field, described by the Stark effect. 

By neglecting the nuclear acceleration term $m\ddot{\mathbf{R}}_s(t)\cdot \hat{\mathbf{r}}$ in the Hamiltonian $ \hat{H}'^{\mathrm{AE}}(\hat{\mathbf{r}}, \hat{\mathbf{p}};\ddot{\mathbf{R}}_s(t))$, $\ket{\phi'_{si}(t)}=\ket{\phi^{\rm \mathbf{0}}_{si}}$, implying that $\ket{\phi'_{si}(t)}$ does not depend on the history of the system and that the atomic orbitals for static nuclei $\ket{\phi^{\rm \mathbf{0}}_{si}}$ satisfy Eq. \eqref{eq:Schrodinger_phi0_moving_ions}.  

Therefore, for moving nuclei the atomic orbitals should be expressed in the form of Eq. \eqref{eq:atomic_orbital_I}. The purely time-dependent phases, including those depending on the history of the system, do not affect the basis and can be removed. Consequently, we define the velocity-including localised atomic orbital  as
\begin{align}
\ket{\phi^{\dot{\mathbf{R}}_s,\mathbf{R}_{s}}_{si}}=e^{i\frac{m}{\hbar}\dot{\mathbf{R}}_s(t)\cdot(\hat{\mathbf{r}}-\mathbf{R}_s(t))}\ket{\phi^{\mathbf{R}_{s}(t)}_{si}}.\label{eq:vi_atomic_orbital}
\end{align} 
In the superscript $\dot{\mathbf{R}}_s,\mathbf{R}_{s}$ we omit the temporal dependence for brevity since it is clear from the context. The velocity-including basis set $\{{\ket{\phi^{\dot{\mathbf{R}}_s,\mathbf{R}_{s}}_{si}}}\}_{si}$ depend on time instantaneously through the nuclear position $\mathbf{R}_s(t)$ and velocity $\dot{\mathbf{R}}_s(t)$. 
The temporal derivative of the states is 
\begin{equation}
    \begin{aligned}
&\frac{d\ket{\phi^{\dot{\mathbf{R}}_s,\mathbf{R}_{s}}_{si}}}{dt}=\frac{i}{\hbar}e^{i\alpha_s(\hat{\mathbf{r}})}\Bigg(m\ddot{\mathbf{R}}_s(t)\cdot(\hat{\mathbf{r}}-\mathbf{R}_s(t))\\&-\dot{\mathbf{R}}_s(t)\cdot \hat{\mathbf{p}}-m|\dot{\mathbf{R}}_s(t)|^2\Bigg) 
\ket{\phi^{\mathbf{R}_{s}(t)}_{si}},\label{eq:derivative_vi_atomic_orbital}
\end{aligned} 
\end{equation}
and, by using that $e^{i\alpha_{s}(\hat{\mathbf{r}})}\hat{\mathbf{p}}e^{-i\alpha_{s}(\hat{\mathbf{r}})}=\hat{\mathbf{p}}-m\dot{\mathbf{R}}_s(t)$, 
\begin{equation}
    \begin{aligned}
\frac{d\ket{\phi^{\dot{\mathbf{R}}_s,\mathbf{R}_{s}}_{si}}}{dt}=&\frac{i}{\hbar}\Bigg(m\ddot{\mathbf{R}}_s(t)\cdot(\hat{\mathbf{r}}-\mathbf{R}_s(t))\\&-\dot{\mathbf{R}}_s(t)\cdot \hat{\mathbf{p}}\Bigg) 
\ket{\phi^{\dot{\mathbf{R}}_s,\mathbf{R}_{s}}_{si}}.\label{eq:derivative_vi_atomic_orbital_v2}
\end{aligned} 
\end{equation}
In the construction of the localised atomic orbital basis set, neglecting the nuclear acceleration term, which induces a Stark effect, causes a small error.  Indeed, we are excluding the mixing, due to the Stark effect, of the occupied states with orbitals that are not included in the basis set. These contributions are small because of the large energy difference between the occupied states and those excluded from the basis set, becoming increasingly smaller as the basis set is enlarged. 

The notation used to indicate the atomic orbitals is summarised in Table \ref{tab:atomic_orbitals_legend}.
From now on, we omit the $\hat{}$ unless it is important for context. 
\begin{table}[h!]
    \centering
    \renewcommand{\arraystretch}{2} % Aumenta lo spazio tra le righe
    \begin{tabular}{l|l}
    &  Atomic orbital for level $i$ of atom $s$\\
    \hline
    $\ket{\phi^{\mathbf{0}}_{si}}$   
        & for the atom centred at the origin \\
    \hline
    $\ket{\phi^{\mathbf{R}_{s}}_{si}}$   
        & translated to the fixed atomic equilibrium position \\
    \hline
    $\ket{\phi^{\mathbf{R}_{s}(t)}_{si}}$   
        & translated to the time-dependent atomic position \\
    \hline
    $\ket{\phi^{\dot{\mathbf{R}}_s,\mathbf{R}_{s}}_{si}}$   
        & $\ket{\phi^{\mathbf{R}_{s}(t)}_{si}}$ times nuclear velocity-dependent phase \\
    \end{tabular}
    \caption{Summary of the notation used for the different kinds of atomic orbitals throughout the paper.}
    \label{tab:atomic_orbitals_legend}
\end{table}

\section{Non-Adiabatic Ehrenfest dynamics}
\label{sec:II}
Consider a system of classical nuclei with mass $M_s$, located at the positions $\{\mathbf{R}_s(t)\}$, and quantum electrons.  The all-electron single-particle mean-field Hamiltonian $\hat{H}$ presents an effective self-consistent local potential $V(\hat{\mathbf{r}})$, obtained with a density functional theory (DFT) local or semi-local approximation for the exchange-correlation functional.  We do not explicitly treat the self-consistent potential since it is local, adding no complications compared with the non-self-consistent case for our purposes. Therefore, the single-particle Hamiltonian of the electron interacting with many nuclei is
\begin{equation}
\hat{H}^{\mathrm{AE}}(\hat{\mathbf{r}},\hat{\mathbf{p}};\{\mathbf{R}_s(t)\})=\frac{\hat{\mathbf{p}}^2}{2m}+V(\hat{\mathbf{r}}).
\label{eq:HAEsingle}
\end{equation}
At zero temperature, the system of electrons and nuclei can be described through the real-valued Ehrenfest Lagrangian \cite{Todorov_2001}, depending on the independent variables $\mathbf{q}=\left(\{\mathbf{R}_s(t)\}_s, \{\ket{\psi_I(t)}\}_I, \{\bra{\psi_I(t)}\}_I\right)$,
\begin{equation}
\begin{aligned}
    & \mathcal{L}(\mathbf{q})=\sum_s\frac{M_s|\dot{\mathbf{R}}_s(t)|^2}{2}-\sum_{I=1}^{N_{\rm el}}\Bigg(\braket{\psi_I(t)|\hat{H}^{\mathrm{AE}}|\psi_I(t)}\\&-\frac{i\hbar}{2}\left(\bra{\psi_I(t)} \frac{d\ket{\psi_I(t)}}{dt}- \frac{d\bra{\psi_I(t)}}{dt}\ket{\psi_I(t)}\right)\Bigg)
\end{aligned}
    \label{eq:lagrangian_general}
\end{equation} 
where $\ket{\psi_I(t)}$ are the solutions of Eq. \eqref{eq:HAEsingle}, and $N_{\rm el}$ is the number of electrons. In addition, if we apply a transformation that depends on time to the wavefunction 
\begin{equation}
    \ket{\psi'(t)}=\mathcal{R}(t)\ket{\psi(t)},
\end{equation}
the Ehrenfest Lagrangian of Eq. \eqref{eq:lagrangian_general} is transformed to
\begin{widetext}
    \begin{equation}
    \begin{aligned}
\mathcal{L}'\left(\mathbf{q}'\right)=\sum_s\frac{M_s|\dot{\mathbf{R}}_s(t)|^2}{2}-&\sum_{I=1}^{N_{\rm el}}\Bigg[\braket{\psi'_I(t)|\mathcal{R}^{\dagger}(t)\hat{H}^{\rm AE}\mathcal{R}(t)|\psi'_I(t)}+\frac{i\hbar}{2}\Bigg(\braket{\psi'_I(t)| \frac{d\mathcal{R}^{\dagger}(t)}{dt}\mathcal{R}(t)-\mathcal{R}^{\dagger}(t)\frac{d\mathcal{R}(t)}{dt}| \psi'_I(t)}\\&+\frac{d\bra{\psi'_I(t)}}{dt}\mathcal{R}^{\dagger}(t)\mathcal{R}(t)\ket{\psi'_I(t)}-\bra{\psi'_I(t)}\mathcal{R}^{\dagger}(t)\mathcal{R}(t)| \frac{d\ket{\psi'_I(t)}}{dt}\Bigg)\Bigg],
\end{aligned}
\label{eq:transformed_lagrangian_general}
\end{equation}
\end{widetext}
where the independent variables are $\mathbf{q}'=\left(\{\mathbf{R}_s(t)\}_s, \{\ket{\psi'_I(t)}\}_I, \{\bra{\psi'_I(t)}\}_I\right)$. We do not treat the finite temperature case directly in the Lagrangian, but it can be obtained with straightforward generalisations. In the following sections, we use Eq. \eqref{eq:transformed_lagrangian_general} to determine the Ehrenfest Lagrangian for the Projector Augmented Wave (PAW) method with the inclusion of the nuclear velocity phases in the atomic orbitals.

The equations of motion for the nuclear and electronic variables are determined by imposing the action $\mathcal{A}=1/T\int_0^T \mathcal{L}dt$ to be stationary for a variation of an independent variable of the Lagrangian, i.e.  $\frac{\delta \mathcal{A}}{\delta q_i}=0$.  Usually, the Lagrangian depends on $\mathbf{q} \textrm{ and } \dot{\mathbf{q}}$, implying that the equations of motion are the Euler-Lagrange equations 
\begin{equation}
    \frac{\partial \mathcal{L}(\mathbf{q},\dot{\mathbf{q}})}{\partial q_i}-\frac{d}{dt}\frac{\partial \mathcal{L}(\mathbf{q},\dot{\mathbf{q}})}{\partial \dot{q}_i}=0,
    \label{eq:Euler-Lagrange}
\end{equation}
which are always used in this work for the electronic degrees of freedom (wavefunctions). The conserved energy is obtained as the Legendre transformation of the Lagrangian 
 \begin{equation}
     E=\sum_{i}\frac{\partial \mathcal{L}(\mathbf{q},\dot{\mathbf{q}})}{\partial \dot{q}_i}\dot{q}_i-\mathcal{L}(\mathbf{q},\dot{\mathbf{q}}).
     \label{eq:conserved_energy}
 \end{equation}
Conversely, the nuclear-velocity dependence of the atomic orbitals may lead to a nuclear-acceleration dependence in the effective model's Lagrangian. Consequently, the Euler-Lagrange equations need to be generalised to higher orders to describe nuclear dynamics. Under the assumption that the highest order temporal derivative is the second order one, the Euler-Lagrange equations are generalised as \cite{woodard2015theoremostrogradsky, ostrogradsky1850memoires, PhysRevD.91.085009}
\begin{equation}
    \frac{\partial \mathcal{L}(\mathbf{q},\dot{\mathbf{q}},\ddot{\mathbf{q}})}{\partial q_i}-\frac{d}{dt}\frac{\partial \mathcal{L}(\mathbf{q},\dot{\mathbf{q}},\ddot{\mathbf{q}})}{\partial \dot{q}_i}+\frac{d^2}{dt^2}\frac{\partial \mathcal{L}(\mathbf{q},\dot{\mathbf{q}},\ddot{\mathbf{q}})}{\partial \ddot{q}_i}=0.
    \label{eq:Euler-Lagrange_general}
\end{equation}
Specifically, the equations for the nuclear system are obtained by setting $q_i \to \mathbf{R}_s$. Explicitly, if the Lagrangian depends also on the nuclear acceleration, the forces governing the nuclear dynamics are 
\begin{equation}
    M_s \ddot{\mathbf{R}}_s(t)=\frac{\partial \mathcal{L}_{\mathrm{el}}}{\partial \mathbf{R}_s}-\frac{d}{dt}\frac{\partial \mathcal{L}_{\mathrm{el}}}{\partial \dot{\mathbf{R}}_s}+\frac{d^2}{dt^2}\frac{\partial \mathcal{L}_{\mathrm{el}}}{\partial \ddot{\mathbf{R}}_s}
    \label{eq:forces_general}
\end{equation}
where $\displaystyle \mathcal{L}_{\mathrm{el}}=\mathcal{L}-\sum_s\frac{M_s|\dot{\mathbf{R}}_s(t)|^2}{2}$.  If Lagrangian depends linearly on the acceleration, the higher order derivative in the equations of motion is the second order one, avoiding issues of Ostrogradsky’s instabilities. In the presence of a dependence on second order derivatives of the variables $\mathbf{q}$ in the Lagrangian, the energy is obtained through Ostrogradsky’s construction \cite{ostrogradsky1850memoires, woodard2015theoremostrogradsky,PhysRevD.91.085009} as 
 \begin{align}
     E=\sum_{i}\frac{\partial \mathcal{L}(\mathbf{q},\dot{\mathbf{q}},\ddot{\mathbf{q}})}{\partial \ddot{q}_i}\ddot{q}_i-\mathcal{L}(\mathbf{q},\dot{\mathbf{q}},\ddot{\mathbf{q}})\nonumber\\
     +\sum_{i}\Bigg[\left(\frac{\partial \mathcal{L}(\mathbf{q},\dot{\mathbf{q}},\ddot{\mathbf{q}})}{\partial \dot{q}_i}-\frac{d}{dt}\frac{\partial \mathcal{L}(\mathbf{q},\dot{\mathbf{q}},\ddot{\mathbf{q}})}{\partial \ddot{q}_i}\right)\dot{q}_i\Bigg].
     \label{eq:conserved_energy_Ostrograsky}
 \end{align}
 As for the equations of motion, the linearity of the Lagrangian on $\ddot{\mathbf{q}}$ forbids any derivatives of order higher than the second in the conserved energy. 

In the absence of an explicit time-dependence in the Lagrangian, the total energy is conserved because the time-derivative of $E$ is zero along the trajectory of the system. All the Lagrangians presented in this work do not have an explicit time dependence. This implies that the energy, either obtained in the usual Hamiltonian formulation of Eq. \eqref{eq:conserved_energy} or with the Ostrogradsky generalisation of Eq. \eqref{eq:conserved_energy_Ostrograsky}, is always conserved.

\section{PAW Method}\label{sec:PAW}
The Projector Augmented Wave (PAW) method, developed in Ref. \cite{PhysRevB.50.17953}, determines the single-particle pseudo-wavefunction $\ket{\tilde{\psi}}$ through a linear transformation $\hat{\mathcal{T}}$ applied to the all-electron wavefunction
\begin{equation}
\ket{\psi}=\hat{\mathcal{T}}\ket{\tilde{\psi}}.
\end{equation} 
In the original formulation, using atomic orbitals for nuclei at rest centred at the positions of the nuclei, the linear transformation is defined by selecting a set of all-electron partial waves $\ket{\phi^{\mathbf{R}_s}_{si}}$ obtained by applying the transformation $\hat{\mathcal{T}}$ to the set of pseudo-partial waves $\ket{\tilde{\phi}^{\mathbf{R}_s}_{si}}$
\begin{equation}
\begin{aligned}
\hat{\mathcal{T}}=\hat{\mathbb{1}}+\sum_{s}\hat{t}^s,\\
\hat{t}^s=\sum_i\left(\ket{\phi^{\mathbf{R}_s}_{si}}- \ket{\tilde{\phi}^{\mathbf{R}_s}_{si}}\right)\bra{\tilde{p}^{\mathbf{R}_s}_{si}}
\end{aligned}
\label{eq:PAW_transformation}
\end{equation}
where the projectors $\ket{\tilde{p}^{\mathbf{R}_s}_{si}}$ satisfy the orthogonality relation with the pseudo-partial waves $\braket{\tilde{p}^{\mathbf{R}_s}_{si}|\tilde{\phi}^{\mathbf{R}_{s'}}_{s'j}}=\delta_{ss'}\delta_{ij}$. We further assume that there exists an augmentation region $\Omega_s$ around each site $\mathbf{R}_s$, where the all-electron partial waves $\ket{\phi^{\mathbf{R}_s}_{si}}$ form a complete set for the valence all-electron wavefunction. Outside the region, the all-electron and pseudo-partial waves coincide and the pseudo-projectors vanish. Finally, the augmentation regions do not overlap. For a complete basis set, if $\mathbf{r} \in \Omega_s$, the following relations hold
\begin{equation}
\begin{aligned}
\sum_{i}\braket{\mathbf{r}|\tilde{\phi}^{\mathbf{R}_{s}}_{si}}\braket{\tilde{p}^{\mathbf{R}_s}_{si}|\tilde{\psi}}=\braket{\mathbf{r}|\tilde{\psi}}, \\
\sum_{i}\braket{\tilde{\psi}|\tilde{p}^{\mathbf{R}_{s}}_{si}}\braket{\tilde{\phi}^{\mathbf{R}_s}_{si}|\mathbf{r}}=\braket{\tilde{\psi}|\mathbf{r}}.
\end{aligned}
\label{eq:identity_augmentation_region}
\end{equation} 

Any local or semi-local operator transforms as \cite{PhysRevB.50.17953}
\begin{align}
\tilde{O}=\hat{\mathcal{T}}^{\dag}O\hat{\mathcal{T}}=
O+\sum_{s,ij} \ket{\tilde{p}^{\mathbf{R}_s}_{si}} \Delta O^s_{ij}\bra{\tilde{p}^{\mathbf{R}_s}_{sj}},
\label{eq:PAW_operator_transformation}\\
\Delta O^s_{ij}=\braket{\phi^{\mathbf{R}_s}_{si}|O|\phi^{\mathbf{R}_s}_{sj}}-\braket{\tilde{\phi}^{\mathbf{R}_s}_{si}|O|\tilde{\phi}^{\mathbf{R}_s}_{sj}},
\label{eq:delta_O}
\end{align}
where the second equality requires the use of the identity in the augmentation region as given in Eq. \eqref{eq:identity_augmentation_region}, which holds for a complete basis set. The transformation rule for the operators of Eq. \eqref{eq:PAW_operator_transformation} implies that the identity operator in the all-electron wavefunction space transforms as 
\begin{equation}
\begin{aligned}
S=\hat{\mathcal{T}}^{\dagger}\hat{\mathcal{T}}=&\mathbb{1}+\sum_{s,ij}\ket{\tilde{p}^{\mathbf{R}_s}_{si}}Q^s_{ij}\bra{\tilde{p}^{\mathbf{R}_s}_{sj}}\\
Q^s_{ij}=&\braket{\phi^{\mathbf{R}_s}_{si}|\phi^{\mathbf{R}_s}_{sj}}-\braket{\tilde{\phi}^{\mathbf{R}_s}_{si}|\tilde{\phi}^{\mathbf{R}_s}_{sj}},
\end{aligned}
\label{eq:TdaggerT_identity}
\end{equation}
where $\mathbb{1}$ indicates the identity in the pseudo-wavefunction space. As a consequence, the pseudo-wavefunction are normalised as $\braket{\psi|S|\psi}=1$. In the norm-conserving case, where $\braket{\phi^{\mathbf{R}_s}_{si}|\phi^{\mathbf{R}_s}_{sj}}=\braket{\tilde{\phi}^{\mathbf{R}_s}_{si}|\tilde{\phi}^{\mathbf{R}_s}_{sj}}$, the identity in the all-electron wave-function space is mapped into the identity in the pseudo-wavefunction space $S=\mathbb{1}$. The PAW Hamiltonian is analogously obtained as
\begin{equation}
\begin{aligned}
    \hat{H}^{\rm PAW}=&\hat{\mathcal{T}}^{\dag}\hat{H}^{\rm AE}\hat{\mathcal{T}}
   = \frac{\mathbf{p}^2}{2m}+V^{\mathrm{loc}}+\sum_{s}v_{s}^{\mathrm{nl}}.
\end{aligned}
\label{eq:PAW_Hamiltoniana_standard}
\end{equation}
$V^{\mathrm{loc}}$ is the local part of the pseudo-potential, plus eventually a DFT self-consistent potential; $v_{s}^{\mathrm{nl}}$ is the non-local part of the potential obtained as in Ref.  \cite{PhysRevB.50.17953}
\subsection{Adiabatic PAW dynamics}
In the adiabatic limit, the electronic wavefunction remains in the ground state, satisfying the eigenvalue equation \cite{PhysRevB.50.17953, PhysRevB.41.7892, PhysRevB.59.1758, rostgaard2009projectoraugmentedwavemethod}, where $H^{\mathrm{PAW}}$ and $S$ are time-dependent through the nuclear motion as well as the instantaneous eigenvalues and eigenvectors,
\begin{equation}
H^{\mathrm{PAW}}\ket{\tilde{\psi}_I}=E_IS\ket{\tilde{\psi}_I}
\end{equation}
\begin{comment}
    The solution to the Schrödinger equation
\begin{equation}
    i\hbar S\frac{d\ket{\tilde{\psi}_I}}{dt}=H^{\mathrm{PAW}}\ket{\tilde{\psi}_I}
    \label{eq:Schr_adiabatic}
\end{equation}
is the trivial quantum-mechanical temporal evolution of the state 
\begin{equation}
    \ket{\tilde{\psi}_I(t)}=e^{-\frac{i}{\hbar}E_It}\ket{\tilde{\psi}_I(t=0)}.
\end{equation}

\end{comment}
The forces on the ions are \cite{PhysRevB.47.10142, PhysRevB.59.1758, PhysRevB.56.R11369, PhysRevB.64.235118}
\begin{equation}
M_s\ddot{\mathbf{R}}_s(t)=\mathbf{F}_s^{\mathrm{HF}}+\mathbf{F}_s^{\mathrm{S}}
\end{equation}    
\begin{equation}
    \begin{aligned}
        &\mathbf{F}_s^{\mathrm{HF}}=-\sum_{I=1}^{N_{\rm el}}\braket{\tilde{\psi}_I|\frac{\partial H^{\mathrm{PAW}}}{\partial \mathbf{R}_s}|\tilde{\psi}_I},
        \end{aligned}
    \label{eq:HF_force_adiabatic}
\end{equation}
\begin{equation}
    \begin{aligned}
        &\mathbf{F}_s^{S}=\sum_{I=1}^{N_{\rm el}}E_I\bra{\tilde{\psi}_I}\frac{\partial S}{\partial \mathbf{R}_s}\ket{\tilde{\psi}_I}.
    \end{aligned}
    \label{eq:S_force_adiabatic}
\end{equation}
where the first contribution is the standard Hellmann-Feynman force $\mathbf{F}_s^{\mathrm{HF}}$, i.e. the quantum mechanical average over the electronic ground state of the variation of the electronic Hamiltonian due to the nuclear displacement; the second one $\mathbf{F}_s^{S}$ is the force originating from the transformation of the identity operator in the pseudo-wavefunction space. In the case of a norm-conserving pseudo-potential $\mathbf{F}_s^{S}$ vanishes, leaving only the  $\mathbf{F}_s^{\mathrm{HF}}$ contribution to the forces. 

\subsection{Standard Non-adiabatic PAW Ehrenfest dynamics}\label{subsec:dynamics_PAW_standard}
In this section, we present the equations of the PAW non-adiabatic Ehrenfest dynamics obtained with the electronic wavefunction rigidly shifting with the nuclei. These equations are presented in Refs. \cite{10.1063/1.3700800, 10.1063/5.0252559, PhysRevB.73.035408} and currently implemented in the GPAW code \cite{10.1063/5.0182685}.  

The Ehrenfest Lagrangian is obtained using Eq. \eqref{eq:transformed_lagrangian_general} with the transformation $\hat{\mathcal{T}}$, defined in Eq. \eqref{eq:PAW_transformation}, with the nuclear positions explicitly depending on time $\mathbf{R}_s(t)$,
    \begin{equation}
    \begin{aligned}
    &\mathcal{L}^{\mathrm{PAW}}_{\rm R}=\sum_s\frac{M_s|\dot{\mathbf{R}}_s(t)|^2}{2}-\sum_{I=1}^{N_{\rm el}}\Bigg[\braket{\tilde{\psi}_I(t)|\hat{H}^{\mathrm{PAW}}_{\rm R}|\tilde{\psi}_I(t)}\\&+\frac{i\hbar}{2}\left(\frac{d\bra{\tilde{\psi}_I(t)}}{dt}S\ket{\tilde{\psi}_I(t)}-\bra{\tilde{\psi}_I(t)}S| \frac{d\ket{\tilde{\psi}_I(t)}}{dt}\right)\Bigg],
\end{aligned}
\label{eq:PAW_lagrangian_NOVI}
\end{equation}
where
\begin{align}
    \hat{H}^{\mathrm{PAW}}_{\rm R}&=\hat{H}^{\rm PAW}+\frac{i\hbar}{2} \left(\frac{d\hat{\mathcal{T}}^{\dagger}}{dt}\hat{\mathcal{T}}-\hat{\mathcal{T}}^{\dagger}\frac{d\hat{\mathcal{T}}}{dt}\right).
    \label{eq:paw_hamiltonian_R}
\end{align}
The $S$ matrix is here time dependent, differently from the adiabatic case, due to the time dependence of the nuclear positions $\mathbf{R}_s(t)$ to be used inside Eq. \eqref{eq:TdaggerT_identity}. As shown in detail in Appendix \ref{app:computation}, the temporal derivative of the transformation, entering in Eq. \eqref{eq:paw_hamiltonian_R}, is \cite{PhysRevB.73.035408,10.1063/1.3700800}
\begin{equation}
    \frac{d\hat{\mathcal{T}}}{dt}=- \frac{i}{\hbar} \sum_s  \left[\dot{\mathbf{R}}_s(t)\cdot \mathbf{\hat{p}},\hat{t}^s\right],
    \label{eq:derivative_T_velocity_including}
\end{equation}
implying that 
\begin{equation}
\begin{aligned}
&        \frac{i\hbar}{2} \left(\frac{d\hat{\mathcal{T}}^{\dagger}}{dt}\hat{\mathcal{T}}-\hat{\mathcal{T}}^{\dagger}\frac{d\hat{\mathcal{T}}}{dt}\right)=-\sum_{s}\dot{\mathbf{R}}_s(t) \cdot \mathbf{C}_{s},  \\&\mathbf{C}_{s}=\sum_{ij}\Bigg(-\left\{\frac{\hat{\mathbf{p}}}{2},\ket{\tilde{p}^{\mathbf{R}_s(t)}_{si}}Q^{s}_{ij}\bra{\tilde{p}^{\mathbf{R}_s(t)}_{sj}} \right\}\\
        &+\ket{\tilde{p}^{\mathbf{R}_s(t)}_{si}} \Delta \hat{\mathbf{p}}^s_{ij}\bra{\tilde{p}^{\mathbf{R}_s(t)}_{sj}}\Bigg),
        \label{eq:M_no_vel}
\end{aligned}
\end{equation}
where $\{,\}$ is the anti-commutator, and $\Delta$ is used in the sense of Eq. \eqref{eq:delta_O}. 
Therefore, the electronic and nuclear dynamics are obtained from the standard Euler-Lagrange equations (Eq. \eqref{eq:Euler-Lagrange}), 
\begin{equation}
    i\hbar S\frac{d\ket{\tilde{\psi}_I(t)}}{dt}=\left(H^{\mathrm{PAW}}_{\rm R}-\frac{1}{2}\frac{dS}{dt}\right) \ket{\tilde{\psi}_I(t)},
    \label{eq:Schrodinger_PAW_no_vel}
\end{equation}
where the effective Hamiltonian governing the electronic dynamics is
\begin{equation}
    \begin{aligned}
        H^{\mathrm{PAW}}_{\rm R}-\frac{1}{2}\frac{dS(t)}{dt}=H^{\mathrm{PAW}}-i\hbar \hat{\mathcal{T}}^{\dagger}\frac{d\hat{\mathcal{T}}}{dt},\\
      -i\hbar \hat{\mathcal{T}}^{\dagger}\frac{d\hat{\mathcal{T}}}{dt}=  -\sum_s \dot{\mathbf{R}}_s(t)\cdot \Bigg(\ket{\tilde{p}^{\mathbf{R}_s(t)}_{si}} \Delta \hat{\mathbf{p}}^s_{ij}\bra{\tilde{p}^{\mathbf{R}_s(t)}_{sj}}\\-\ket{\tilde{p}^{\mathbf{R}_s(t)}_{si}}Q^{s}_{ij}\bra{\tilde{p}^{\mathbf{R}_s(t)}_{sj}}\frac{\hat{\mathbf{p}}}{2} \Bigg).
    \end{aligned}
\end{equation}
Compared to the adiabatic dynamics, there is the additional time-dependent term $\displaystyle -i\hbar \hat{\mathcal{T}}^{\dagger}\frac{d\hat{\mathcal{T}}}{dt}$.  This term introduces coupling between different electronic states, even when the entire system undergoes a global translation at constant speed, breaking Galilean invariance \cite{doi:10.1021/jz3006173}. As shown in Appendix \ref{app:temporal}, the presence of time-dependent $S$ makes it so the evolution of the pseudo-wavefunction is determined by a non-hermitian operator $\displaystyle H^{\mathrm{PAW}}_{\rm R}-\frac{1}{2}\frac{dS}{dt}$. This allows for the norm conservation of $\displaystyle \braket{\tilde{\psi}(t)|S|\tilde{\psi}(t)}$ along the temporal evolution.

%The use of real-valued Lagrangian allows to have a manifestly Hermitian $P$ operator, at odd with the non-hermitian expressions of Refs. \cite{PhysRevB.73.035408, 10.1063/1.3700800, 10.1063/5.0182685, 10.1063/5.0252559}. 
According to Eq. \eqref{eq:forces_general}, the nuclear dynamics is governed by the following equation

\begin{equation}
M_s\ddot{\mathbf{R}}_s(t)=\mathbf{F}_s^{\mathrm{HF}}+\mathbf{F}_s^{\mathrm{HF-\dot{R}}}+\mathbf{F}_s^{\mathrm{S}}
    \label{eq:Ehrenfest_ions_no_vel}
\end{equation}
where 
\begin{equation}
    \begin{aligned}
        \mathbf{F}_s^{\mathrm{HF}}=&-\sum_{I=1}^{N_{\rm el}}\braket{\tilde{\psi}_I(t)|\frac{\partial H^{\mathrm{PAW}}_{\rm R}}{\partial \mathbf{R}_s}|\tilde{\psi}_I(t)},\\
        \mathbf{F}_s^{\mathrm{HF-\dot{R}}}=&\sum_{I=1}^{N_{\rm el}}\frac{d}{dt}\braket{\tilde{\psi}_I(t)|\frac{\partial H^{\mathrm{PAW}}_{\rm R}}{\partial \dot{\mathbf{R}}_s}|\tilde{\psi}_I(t)},\\
       & \text{with  }\\ &\frac{\partial H^{\mathrm{PAW}}_{\rm R}}{\partial \dot{\mathbf{R}}_s}=-\mathbf{C}_s,\\
        \end{aligned}
    \label{eq:HF_force_pseudo_novel}
\end{equation}

\begin{equation}
    \begin{aligned}
        \mathbf{F}_s^{S}=&\frac{i\hbar}{2}\sum_{I=1}^{N_{\rm el}}\Bigg(\bra{\tilde{\psi}_I(t)}\frac{\partial S}{\partial \mathbf{R}_s}\frac{d \ket{\tilde{\psi}_I(t)}}{dt}\\
        & -\frac{d\bra{\tilde{\psi}_I(t)}}{dt}\frac{\partial S}{\partial \mathbf{R}_s}\ket{\tilde{\psi}_I(t)}\Bigg).
    \end{aligned}
     \label{eq:HF_force_pseudo_novel_S}
\end{equation}
$\mathbf{F}_s^{\mathrm{HF}}$ and $\mathbf{F}_s^{S}$ reduce to Eqs. \eqref{eq:HF_force_adiabatic} and \eqref{eq:S_force_adiabatic} in the adiabatic limit where $\displaystyle H^{\mathrm{PAW}}_{\rm R}$ reduces to $\displaystyle H^{\mathrm{PAW}}$.  In addition, in the non-adiabatic dynamics, an additional Hellmann-Feynman contribution $\mathbf{F}_s^{\mathrm{HF-\dot{R}}}$ appears due to the nuclear velocity dependence in the electronic part of the Lagrangian, expressed as the time derivative of a Hellmann-Feynman term entailing the nuclear velocity derivative \cite{10.1063/1.3700800}. 

Finally, according to Eq. \eqref{eq:conserved_energy}, the conserved energy is
\begin{equation}
    \begin{aligned}
        E_{\rm R}^{\mathrm{PAW}}=\sum_s\frac{M_s}{2}|\dot{\mathbf{R}}_s(t)|^2+\sum_{I=1}^{N_{\rm el}}\braket{\tilde{\psi}_I(t)|\hat{H}^{\rm PAW}|\tilde{\psi}_I(t)},
    \end{aligned}
    \label{eq:conserved_energy_PAW}
\end{equation}
where all the terms that are linear in the derivatives of the independent variables of the Lagrangian cancel.
\begin{comment}
    The nuclear momentum is  
\begin{equation}
  \mathbf{P}=\sum_s \left( M_s\dot{\mathbf{R}}_s(t)-\sum_{I=1}^{N_{\rm el}}\braket{\tilde{\psi}_I(t)|\frac{\partial \mathcal{M}}{\partial \dot{\mathbf{R}}_s}|\tilde{\psi}_I(t)} \right)
  \label{eq:momentum_NOVI}
\end{equation}
where 
\begin{equation}
    \frac{\partial \mathcal{M}}{\partial \dot{\mathbf{R}}_s}=-\mathbf{C}_{s}.
\end{equation}

\end{comment}

\subsubsection*{Norm-conserving case}
In the norm-conserving case, where $S=\mathbb{1}$, $Q^s_{ij}=0$ and
\begin{equation}
\begin{aligned}
\mathbf{C}^{\rm NC}_{s}=\sum_{ij}\ket{\tilde{p}^{\mathbf{R}_s(t)}_{si}} \Delta \hat{\mathbf{p}}^s_{ij}\bra{\tilde{p}^{\mathbf{R}_s(t)}_{sj}},
        \label{eq:M_no_vel_NC}
\end{aligned}
\end{equation}
the Hamiltonian reduces to 
\begin{equation}
  H^{\mathrm{PAW-NC}}_{\mathrm{R}}=H^{\mathrm{PAW}}-\sum_{s}\dot{\mathbf{R}}_s(t) \cdot \mathbf{C}^{\rm NC}_{s}.
  \label{eq:paw_hamiltonian_R_NC}
\end{equation}
The Schrödinger equation for the pseudo-wavefunctions is 
\begin{equation}
    i\hbar \frac{d\ket{\tilde{\psi}_I(t)}}{dt}=H^{\mathrm{PAW-NC}}_{\mathrm{R}}\ket{\tilde{\psi}_I(t)}.
    \label{eq:Schrodinger_PAW_no_vel_NC}
\end{equation}
In addition, the nuclear dynamics of Eq. \eqref{eq:Ehrenfest_ions_no_vel} is governed by $\mathbf{F}_s^{\mathrm{HF}}$ and $\mathbf{F}_s^{\mathrm{HF-\dot{R}}}$, obtained from the derivative of the norm-conserving Lagrangian, since $\mathbf{F}_s^{S}$ is zero. 

\section{Velocity-including PAW} \label{sec:PAW_vel}
With an analogous procedure to the one followed in Refs. \cite{PhysRevB.63.245101, PhysRevLett.91.196401, PhysRevB.76.024401} for the inclusion of magnetic fields in the PAW method, we define the nuclear velocity-including PAW transformation operator by substituting the atomic orbitals for fixed nuclei with the velocity-including ones, defined in Eq. \eqref{eq:vi_atomic_orbital}. We assume that the atomic orbitals centred on a nucleus feel only the motion of that nucleus; in other words, atomic orbitals are those obtained as if each nucleus were isolated, and then, we exploit the superposition principle to construct the transformation operator for the entire wavefunction. The pseudo-wavefunction $\ket{\tilde{\psi}}$ is mapped to the velocity-including all-electron one by the transformation operator for velocity-including orbitals $ \hat{\mathcal{T}}_{\dot{\mathrm{R}}}$, 
\begin{align}
&\ket{\psi}=\hat{\mathcal{T}}_{\dot{\mathrm{R}}}\ket{\tilde{\psi}}
\label{eq:pseudo_vi_AE_vi}\\
&\hat{\mathcal{T}}_{\dot{\mathrm{R}}}=\hat{\mathbb{1}}+\sum_{s,i}\hat{t}^s_{\dot{\mathbf{R}}_s}, \label{eq:Trdot}\\
&\hat{t}^s_{\dot{\mathbf{R}}_s}=e^{i\alpha_s(\hat{\mathbf{r}})} 
     \left(\ket{\phi^{\mathbf{R}_s(t)}_{si}}- \ket{\tilde{\phi}^{\mathbf{R}_s(t)}_{si}}\right)\bra{\tilde{p}^{\mathbf{R}_s(t)}_{si}}e^{-i\alpha_s(\hat{\mathbf{r}})}.
    \nonumber
\end{align}
This expression coincides with the one given in the Supplementary of Ref. \cite{PhysRevLett.136.196401}. 
In the following, for brevity we indicate the nuclear velocity phase using $\alpha_{s}(\hat{\mathbf{r}})=\frac{m}{\hbar}\dot{\mathbf{R}}_s(t)\cdot(\hat{\mathbf{r}}-\mathbf{R}_s(t))$, as in Eq. \eqref{eq:atomic_orbital_I}. 

The transformation operator depends on both the time-dependent nuclear positions and velocities.  For a complete basis set, the relations for the standard PAW of Eq. \eqref{eq:identity_augmentation_region}, assuming that $\mathbf{r} \in \Omega_s$, are transformed as 
\begin{equation}
\begin{aligned}
\sum_{i}\bra{\mathbf{r}}\left(e^{i\alpha_s(\hat{\mathbf{r}})}\ket{\tilde{\phi}^{\mathbf{R}_{s}}_{si}}\bra{\tilde{p}^{\mathbf{R}_s}_{si}}e^{-i\alpha_s(\hat{\mathbf{r}})}\right)e^{i\alpha_s(\hat{\mathbf{r}})}\ket{\tilde{\psi}}\\=e^{i\alpha_s(\mathbf{r})}\braket{\mathbf{r}|\tilde{\psi}}, \\
\sum_{i}\bra{\tilde{\psi}}\left(e^{i\alpha_s(\hat{\mathbf{r}})}\ket{\tilde{p}^{\mathbf{R}_{s}}_{si}}\bra{\tilde{\phi}^{\mathbf{R}_s}_{si}}e^{-i\alpha_s(\hat{\mathbf{r}})}\right)e^{i\alpha_s(\hat{\mathbf{r}})}\ket{\mathbf{r}}\\=\braket{\tilde{\psi}|\mathbf{r}} e^{-i\alpha_s(\mathbf{r})}.
\end{aligned}
\label{eq:identity_augmentation_region_vel}
\end{equation} 
The transformation of the operators of Eq. \eqref{eq:PAW_operator_transformation} - also given in the Supplementary of Ref. \cite{PhysRevLett.136.196401} - is generalised to an expression analogous to the case of the presence of a magnetic field \cite{PhysRevB.63.245101}
\begin{equation}
\begin{aligned}
      &\tilde{O}=O+\sum_{s,ij} e^{i\alpha_s(\hat{\mathbf{r}})}\ket{\tilde{p}^{\mathbf{R}_s(t)}_{si}} \Delta O^{s, \mathrm{\dot{R}}}_{ij}\bra{\tilde{p}^{\mathbf{R}_s(t)}_{sj}}e^{-i\alpha_s(\hat{\mathbf{r}})},\\
     &\Delta O^{s, \mathrm{\dot{R}}}_{ij}=\braket{\phi^{\mathbf{R}_s(t)}_{si}|e^{-i\alpha_s(\hat{\mathbf{r}})}Oe^{i\alpha_s(\hat{\mathbf{r}})}|\phi^{\mathbf{R}_s(t)}_{sj}}\\&
     -\braket{\tilde{\phi}^{\mathbf{R}_s(t)}_{si}|e^{-i\alpha_s(\hat{\mathbf{r}})}Oe^{i\alpha_s(\hat{\mathbf{r}})}|\tilde{\phi}^{\mathbf{R}_s(t)}_{sj}}.
     \label{eq:otildevipaw}
\end{aligned}
\end{equation}
The identity operator for the velocity-including case is transformed as the generalisation of the standard expression given in Eq. \eqref{eq:TdaggerT_identity}
\begin{equation}
\begin{aligned}
S_{\rm R, \dot{R}}=\hat{\mathcal{T}}_{\dot{\mathrm{R}}}^{\dagger}\hat{\mathcal{T}}_{\dot{\mathrm{R}}}=&\mathbb{1}+\sum_{s,ij}e^{i\alpha_s(\hat{\mathbf{r}})}\ket{\tilde{p}^{\mathbf{R}_s(t)}_{si}}Q^s_{ij}\bra{\tilde{p}^{\mathbf{R}_s(t)}_{sj}}e^{-i\alpha_s(\hat{\mathbf{r}})},\\
Q^s_{ij}=&\braket{\phi^{\mathbf{R}_s(t)}_{si}|\phi^{\mathbf{R}_s(t)}_{sj}}-\braket{\tilde{\phi}^{\mathbf{R}_s(t)}_{si}|\tilde{\phi}^{\mathbf{R}_s(t)}_{sj}}.
\end{aligned}
\label{eq:TdaggerT_identity_vel}
\end{equation}
$S_{\rm R, \dot{R}}$ depends on both the nuclear positions and velocities. 

\subsection{Velocity-including Ehrenfest dynamics}\label{sec:Ehrnefest_PAW_vel}
In this section, we derive the non-adiabatic Ehrenfest dynamics by using the velocity-including PAW method.
The Ehrenfest Lagrangian is obtained using Eq. \eqref{eq:transformed_lagrangian_general} with the transformation $\hat{\mathcal{T}}_{\dot{\mathrm{R}}}$, defined in Eq. \eqref{eq:Trdot}, as
%\begin{widetext}
 \begin{equation}
    \begin{aligned}
    &\mathcal{L}_{\rm R, \dot{R}}^{\mathrm{PAW}}=\sum_s\frac{M_s|\dot{\mathbf{R}}_s(t)|^2}{2}-\sum_{I=1}^{N_{\rm el}}\Bigg[\braket{\tilde{\psi}_I(t)|\hat{H}^{\rm PAW}_{\mathrm{R,\dot{R}}}|\tilde{\psi}_I(t)}\\&+\frac{i\hbar}{2}\left(\frac{d\bra{\tilde{\psi}_I(t)}}{dt}S_{\rm R, \dot{R}}\ket{\tilde{\psi}_I(t)}-\bra{\tilde{\psi}_I(t)}S_{\rm R, \dot{R}}| \frac{d\ket{\tilde{\psi}_I(t)}}{dt}\right)\Bigg],
\end{aligned}
\label{eq:PAW_lagrangian_VI}
\end{equation}
%\end{widetext}
where 
\begin{equation}
\begin{aligned}
    \hat{H}^{\rm PAW}_{\mathrm{R,\dot{R}}}= &\hat{\mathcal{T}}_{\dot{\mathrm{R}}}^{\dagger}\hat{H}^{\mathrm{AE}}(\hat{\mathbf{r}},\hat{\mathbf{p}};\{\mathbf{R}_s(t)\}) \hat{\mathcal{T}}_{\dot{\mathrm{R}}}+\frac{i\hbar}{2}\left(  \frac{d\hat{\mathcal{T}}_{\dot{\mathrm{R}}}^{\dagger}}{dt}\hat{\mathcal{T}}_{\dot{\mathrm{R}}}-\hat{\mathcal{T}}_{\dot{\mathrm{R}}}^{\dagger}\frac{d\hat{\mathcal{T}}_{\dot{\mathrm{R}}}}{dt}\right).
    \label{eq:HPAWRdotR}
\end{aligned}
\end{equation}
We are going to drop $\{\mathbf{R}_s(t)\}$ in the following. The contribution from the temporal derivative is crucial to obtain the correct Hamiltonian.  By using $e^{-i\alpha_s(\hat{\mathbf{r}})}H^{\mathrm{AE}}(\hat{\mathbf{r}}, \hat{\mathbf{p}})e^{i\alpha_s(\hat{\mathbf{r}})}=H^{\mathrm{AE}}(\hat{\mathbf{r}},\hat{\mathbf{p}}+m\dot{\mathbf{R}}_s(t))$ and the transformation of the operators expressed as in Eq. \eqref{eq:otildevipaw}, it follows that (see App. \ref{app:details_operators}) 
\begin{equation}
\begin{aligned}
    &\hat{\mathcal{T}}_{\dot{\mathrm{R}}}^{\dagger}\hat{H}^{\mathrm{AE}}(\hat{\mathbf{r}},\hat{\mathbf{p}}) \hat{\mathcal{T}}_{\dot{\mathrm{R}}}=\\
    &\frac{\hat{\mathbf{p}}^2}{2m}+V^{\mathrm{loc}}(\hat{\mathbf{r}})+\sum_s e^{i\alpha_s(\hat{\mathbf{r}})} v_s^{\mathrm{nl}}e^{-i\alpha_s(\hat{\mathbf{r}})}
\\
&+\sum_{s,ij} e^{i\alpha_s(\hat{\mathbf{r}})}\ket{\tilde{p}^{\mathbf{R}_s(t)}_{si}} \Delta \mathcal{H}^s_{ij}\bra{\tilde{p}^{\mathbf{R}_s(t)}_{sj}}e^{-i\alpha_s(\hat{\mathbf{r}})},
\end{aligned}
 \label{eq:PAW_Hamiltonian}
\end{equation}
where
\begin{align}
    & \mathcal{H}= H^{\mathrm{AE}}(\hat{\mathbf{r}},\hat{\mathbf{p}}+m\dot{\mathbf{R}}_s(t))-H^{\mathrm{AE}}(\hat{\mathbf{r}},\hat{\mathbf{p}})\\
    &=\dot{\mathbf{R}}_s(t)\cdot \hat{\mathbf{p}}+\frac{m}{2}|\dot{\mathbf{R}}_s(t)|^2, \nonumber
\end{align}

and we stress that $\Delta \mathcal{H}^s_{ij}$ is computed as in Eq. \eqref{eq:delta_O}.
To determine the contribution to the Hamiltonian in Eq. \eqref{eq:HPAWRdotR} originating in the temporal derivatives of the transformation, as expressed in detail in Appendix \ref{app:computation},  we compute $\displaystyle \frac{d\hat{\mathcal{T}}_{\dot{\mathrm{R}}}}{dt}$ that, using Eq. \eqref{eq:derivative_vi_atomic_orbital_v2} and defining 
\begin{equation}
    \mathbf{D}_s(\hat{\mathbf{r}},\hat{\mathbf{p}})=\frac{i}{\hbar}\left(m\mathbf{\ddot{R}}_s(t)\cdot \left(\hat{\mathbf{r}}-\mathbf{R}_s(t)\right)-\dot{\mathbf{R}}_s(t)\cdot \mathbf{\hat{p}}\right),
    \label{eq:orbital_derivative}
\end{equation}
is expressed as
\begin{equation}
    \frac{d\hat{\mathcal{T}}_{\dot{\mathrm{R}}}}{dt}=  \sum_s  \left[\mathbf{D}_s(\hat{\mathbf{r}},\hat{\mathbf{p}}),\hat{t}^s_{\dot{\mathbf{R}}_s}\right].
    \label{eq:derivative_T_velocity_including_dotR}
\end{equation}
$\displaystyle \frac{d\hat{\mathcal{T}}_{\dot{\mathrm{R}}}}{dt}$ vanishes outside each augmentation region $\Omega_s$. As explained in detail in Appendix \ref{app:computation}, we obtain 
\begin{widetext}
\begin{equation}
\begin{aligned}
&\frac{i\hbar}{2}\left(  \frac{d\hat{\mathcal{T}}_{\dot{\mathrm{R}}}^{\dagger}}{dt}\hat{\mathcal{T}}_{\dot{\mathrm{R}}}-\hat{\mathcal{T}}_{\dot{\mathrm{R}}}^{\dagger}\frac{d\hat{\mathcal{T}}_{\dot{\mathrm{R}}}}{dt}\right)= -i\hbar\sum_{s,ij} \Bigg(e^{i\alpha_s(\hat{\mathbf{r}})} \ket{\tilde{p}^{\mathbf{R}_s(t)}_{sj}} \Delta \mathbf{D}^{s,\mathrm{\dot{R}}}_{ij}\bra{\tilde{p}^{\mathbf{R}_s(t)}_{si}}
  e^{-i\alpha_s(\hat{\mathbf{r}})}-\left\{\frac{\mathbf{D}_s}{2},e^{i\alpha_s(\hat{\mathbf{r}})} \ket{p^{\mathbf{R}_s(t)}_{si}}Q^s_{ij}\bra{p^{\mathbf{R}_s(t)}_{sj}}e^{-i\alpha_s(\hat{\mathbf{r}})}\right\}\Bigg).
\end{aligned}
\label{eq:acceleration_PAW_lagrangian}
\end{equation} 
Explicitly:
\begin{equation}
\begin{aligned}
    \Delta \mathbf{D}^{s,\mathrm{\dot{R}}}_{ij}=&\frac{i}{\hbar}\Bigg(\braket{\phi^{\mathbf{R}_s(t)}_{si}|\left[ m\mathbf{\ddot{R}}_s(t)\cdot \left[\hat{\mathbf{r}}-\mathbf{R}_s(t)\right]-\dot{\mathbf{R}}_s(t)\cdot \mathbf{\hat{p}}-m|\dot{\mathbf{R}}_s(t)|^2\right]|\phi^{\mathbf{R}_s(t)}_{sj}}
\\&-\braket{\tilde{\phi}^{\mathbf{R}_s(t)}_{si}|\left[m\mathbf{\ddot{R}}_s(t)\cdot \left[\hat{\mathbf{r}}-\mathbf{R}_s(t)\right]-\dot{\mathbf{R}}_s(t)\cdot \mathbf{\hat{p}}-m|\dot{\mathbf{R}}_s(t)|^2\right]|\tilde{\phi}^{\mathbf{R}_s(t)}_{sj}}\Bigg).
\end{aligned}
\end{equation}
Summing Eqs. \eqref{eq:PAW_Hamiltonian} and \eqref{eq:acceleration_PAW_lagrangian},  we obtain the effective Hamiltonian from Eq. \eqref{eq:HPAWRdotR}
\begin{equation}
\begin{aligned}
&\hat{H}^{\mathrm{PAW}}_{\mathrm{R,\dot{R}}}=  \frac{\hat{\mathbf{p}}^2}{2m}+V^{\mathrm{loc}}(\hat{\mathbf{r}})+\sum_{s}e^{\frac{i}{\hbar}m\dot{\mathbf{R}}_s(t)\cdot \hat{\mathbf{r}}}v_{s}^{\mathrm{nl}}e^{-\frac{i}{\hbar}m\dot{\mathbf{R}}_s(t)\cdot \hat{\mathbf{r}}}
+\sum_{si}\Bigg( m\ddot{\mathbf{R}}_s(t) \cdot e^{i\alpha_s(\hat{\mathbf{r}})} \ket{p^{\mathbf{R}_s(t)}_{si}}\mathbf{d}^s_{ij}\bra{p^{\mathbf{R}_s(t)}_{sj}}e^{-i\alpha_s(\hat{\mathbf{r}})} \\&
    -\frac{1}{2}\left\{m \ddot{\mathbf{R}}_s(t)\cdot\left[\hat{\mathbf{r}}-\mathbf{R}_s(t)\right]-\dot{\mathbf{R}}_s(t)\cdot \hat{\mathbf{p}}+\frac{m|\dot{\mathbf{R}}_s(t)|^2}{2},e^{i\alpha_s(\hat{\mathbf{r}})} \ket{p^{\mathbf{R}_s(t)}_{si}}Q^s_{ij}\bra{p^{\mathbf{R}_s(t)}_{sj}}e^{-i\alpha_s(\hat{\mathbf{r}})} \right\}\Bigg).
\end{aligned}
 \label{eq:PAW_Hamiltonian-Rdot}
\end{equation} 
We have defined, following the notation used in Ref. \cite{PhysRevB.69.235102}, the quantity
\begin{equation}
\begin{aligned}
    \mathbf{d}^{s}_{ij}=&m\mathbf{\ddot{R}}_s(t)\cdot\Bigg(\braket{\phi^{\mathbf{R}_s(t)}_{si}| \left[ \hat{\mathbf{r}}-\mathbf{R}_s(t) \right]|\phi^{\mathbf{R}_s(t)}_{sj}}-\braket{\tilde{\phi}^{\mathbf{R}_s(t)}_{si}|\left[\hat{\mathbf{r}}-\mathbf{R}_s(t)\right] |\tilde{\phi}^{\mathbf{R}_s(t)}_{sj}}\Bigg).
\end{aligned}
\end{equation}
\end{widetext}
The last line of Eq. \eqref{eq:PAW_Hamiltonian-Rdot} are additional terms compared to the nuclear velocity-including pseudo-potential Hamiltonian given in Ref. \cite{PhysRevLett.136.196401}. While vanishing in the norm-conserving case, they can not be neglected for ultra-soft pseudo-potentials, however they do not raise any particular computational issues. Indeed, $\displaystyle \mathbf{d}^s_{ij}$ is easily computed in the ultra-soft pseudo-potential implementations \cite{PhysRevB.64.235118, PhysRevB.69.235102}. Also the term $\displaystyle \left\{\left[\hat{\mathbf{r}}-\mathbf{R}_s(t)\right],e^{i\alpha_s(\hat{\mathbf{r}})}\ket{p^{\mathbf{R}_s(t)}_{si}}Q^s_{ij}\bra{p^{\mathbf{R}_s(t)}_{sj}}e^{-i\alpha_s(\hat{\mathbf{r}})}\right\}$ is well-defined since it is localised in each augmentation region. 

The $\Delta \hat{\mathbf{p}}^s_{ij}$ terms, appearing in Eq. \eqref{eq:M_no_vel}, are cancelled by the momentum translation of the Hamiltonian due to the nuclear velocity-dependent phases. This removes the spurious couplings between electronic states present in Eq. \eqref{eq:Schrodinger_PAW_no_vel}, emerging even when the entire system undergoes a global translation at constant speed. The difference between the two formulations can be appreciated in the comparison of Table \ref{tab:hamiltonians_dynamics}. 

 The electronic dynamics are described by the standard Euler-Lagrange equation (Eq. \eqref{eq:Euler-Lagrange})
\begin{equation}
    i\hbar S\frac{d\ket{\tilde{\psi}_I(t)}}{dt}=\left(\hat{H}^{\mathrm{PAW}}_{\mathrm{R,\dot{R}}} -\frac{1}{2}\frac{dS_{\mathrm{R,\dot{R}}}}{dt}\right)\ket{\tilde{\psi}_I(t)},
    \label{eq:Schrodinger_PAW}
\end{equation}
where 
\begin{equation}
\begin{aligned}
    & \hat{H}^{\mathrm{PAW}}_{\mathrm{R,\dot{R}}} -\frac{1}{2}\frac{dS_{\mathrm{R,\dot{R}}}}{dt}=\\&\hat{\mathcal{T}}_{\dot{\mathrm{R}}}^{\dagger}\hat{H}^{\mathrm{AE}}(\hat{\mathbf{r}},\hat{\mathbf{p}}) \hat{\mathcal{T}}_{\dot{\mathrm{R}}}-i\hbar \hat{\mathcal{T}}_{\dot{\mathrm{R}}}^{\dagger}\frac{d\hat{\mathcal{T}}_{\dot{\mathrm{R}}}}{dt}.
\end{aligned}
\end{equation}
In the presence of nuclear acceleration contributions in the Lagrangian, the nuclear dynamics is governed by Eq. \eqref{eq:forces_general} with the nuclear acceleration terms
 \begin{equation}
M_s\ddot{\mathbf{R}}_s(t)=\mathbf{F}_s^{\mathrm{HF}}+\mathbf{F}_s^{\mathrm{HF-\dot{R}}}+\mathbf{F}_s^{\mathrm{HF-\ddot{R}}}+\mathbf{F}_s^{\mathrm{S}}+\mathbf{F}_s^{\mathrm{S-\dot{R}}}.
    \label{eq:Ehrenfest_ions_1}
\end{equation}
A nuclear acceleration Hellmann-Feynman force $\mathbf{F}_s^{\mathrm{HF-\ddot{R}}}$ appear in the equations for the ions compared to Eq. \eqref{eq:Ehrenfest_ions_no_vel}. The other Hellmann-Feynman terms have analogous expressions to the ones given in Eqs. \eqref{eq:HF_force_pseudo_novel},
\begin{equation}
    \begin{aligned}
        &\mathbf{F}_s^{\mathrm{HF}}=-\sum_{I=1}^{N_{\rm el}}\braket{\tilde{\psi}_I(t)|\frac{\partial \hat{H}^{\mathrm{PAW}}_{\mathrm{R,\dot{R}}}}{\partial \mathbf{R}_s}|\tilde{\psi}_I(t)},\\
        &\mathbf{F}_s^{\mathrm{HF-\dot{R}}}=\sum_{I=1}^{N_{\rm el}}\frac{d}{dt}\braket{\tilde{\psi}_I(t)|\frac{\partial \hat{H}^{\mathrm{PAW}}_{\mathrm{R,\dot{R}}}}{\partial \dot{\mathbf{R}}_s}|\tilde{\psi}_I(t)},\\
        \end{aligned}
    \label{eq:HF_force_pseudo_vel}
\end{equation}
\begin{equation}
    \begin{aligned}
        &\mathbf{F}_s^{\mathrm{HF-\ddot{R}}}=-\sum_{I=1}^{N_{\rm el}}\frac{d^2}{dt^2}\braket{\tilde{\psi}_I(t)|\frac{\partial \hat{H}^{\mathrm{PAW}}_{\mathrm{R,\dot{R}}}}{\partial \ddot{\mathbf{R}}_s}|\tilde{\psi}_I(t)}.
    \end{aligned}
    \label{eq:HF_force_pseudo_vel_bis}
\end{equation}

Crucially, the nuclear velocity derivative entering in $\mathbf{F}_s^{\mathrm{HF-\dot{R}}}$ differs from that of Eq. \eqref{eq:HF_force_pseudo_novel}, being
\begin{align}
   & \frac{\partial \hat{H}^{\mathrm{PAW}}_{\mathrm{R,\dot{R}}}}{\partial \dot{\mathbf{R}}_{s}}=\frac{im}{\hbar}\left[\mathbf{r},v_s^{\mathrm{nl}}\right]+\frac{1}{2}\sum_{ij}\Bigg\{ \hat{\mathbf{p}}-\frac{m\dot{\mathbf{R}}_s(t)}{2},\nonumber\\
&e^{i\alpha_s(\hat{\mathbf{r}})}\ket{p^{\mathbf{R}_s(t)}_{si}}Q^s_{ij}\bra{p^{\mathbf{R}_s(t)}_{sj}}e^{-i\alpha_s(\hat{\mathbf{r}})}\Bigg\}\label{eq:PAW_derivative_dotR}.
\end{align}
The nuclear acceleration derivative is expressed as
\begin{equation}
    \begin{aligned}
    &\frac{\partial \hat{H}^{\mathrm{PAW}}_{\mathrm{R,\dot{R}}}}{\partial \ddot{\mathbf{R}}_{s}}=e^{i\alpha_s(\hat{\mathbf{r}})} \ket{p^{\mathbf{R}_s(t)}_{si}}\mathbf{d}^s_{ij}\bra{p^{\mathbf{R}_s(t)}_{sj}}e^{-i\alpha_s(\hat{\mathbf{r}})}\\
    -&\frac{1}{2}\left\{\left[\hat{\mathbf{r}}-\mathbf{R}_s(t)\right],e^{i\alpha_s(\hat{\mathbf{r}})}\ket{p^{\mathbf{R}_s(t)}_{si}}Q^s_{ij}\bra{p^{\mathbf{R}_s(t)}_{sj}}e^{-i\alpha_s(\hat{\mathbf{r}})}\right\}\Bigg).
    \end{aligned}
    \label{eq:PAW_derivative_ddotR}
\end{equation}
Additional forces originating from $S_{\mathrm{R,\dot{R}}}$  are $\mathbf{F}_s^{S}$ (analogous to Eq. \eqref{eq:HF_force_pseudo_novel_S} and reducing to Eq. \eqref{eq:S_force_adiabatic} in the adiabatic limit) and its nuclear velocity-generalisation $\mathbf{F}_s^{\rm S-\mathrm{\dot{R}}}$, appearing only in the velocity-including PAW,
\begin{equation}
    \begin{aligned}
           \mathbf{F}_s^{\rm S}=&\frac{i\hbar}{2}\sum_{I=1}^{N_{\rm el}}\Bigg(\bra{\tilde{\psi}_I(t)}\frac{\partial S_{\mathrm{R,\dot{R}}}}{\partial \mathbf{R}_s}\frac{d \ket{\tilde{\psi}_I(t)}}{dt}\\
        & -\frac{d\bra{\tilde{\psi}_I(t)}}{dt}\frac{\partial S_{\mathrm{R,\dot{R}}}}{\partial \mathbf{R}_s}\ket{\tilde{\psi}_I(t)}\Bigg),\\      
    \end{aligned}
\end{equation}
\begin{equation}
    \begin{aligned}
 \mathbf{F}_s^{\mathrm{S-\dot{R}}}=&-\frac{i\hbar}{2}\frac{d}{dt}\sum_{I=1}^{N_{\rm el}}\Bigg(\bra{\tilde{\psi}_I(t)}\frac{\partial S_{\mathrm{R,\dot{R}}}}{\partial \dot{\mathbf{R}}_s}\frac{d \ket{\tilde{\psi}_I(t)}}{dt}\\
        & -\frac{d\bra{\tilde{\psi}_I(t)}}{dt}\frac{\partial S_{\mathrm{R,\dot{R}}}}{\partial \dot{\mathbf{R}}_s}\ket{\tilde{\psi}_I(t)}\Bigg),       
    \end{aligned}
\end{equation}
where, from Eq. \eqref{eq:TdaggerT_identity_vel}, the nuclear velocity derivative of $S_{\mathrm{R,\dot{R}}}$ is
\begin{equation}
    \frac{\partial S_{\mathrm{R,\dot{R}}}}{\partial \dot{\mathbf{R}}_s}=\Big[ \hat{\mathbf{r}},e^{i\alpha_s(\hat{\mathbf{r}})} \ket{p^{\mathbf{R}_s(t)}_{si}}Q^s_{ij}\bra{p^{\mathbf{R}_s(t)}_{sj}}e^{-i\alpha_s(\hat{\mathbf{r}})} \Big].
    \label{eq:S_derivative_dotR}
\end{equation}
Finally, the conserved energy is computed according to Eq. \eqref{eq:conserved_energy_Ostrograsky} that holds in the presence of nuclear acceleration terms in the Lagrangian. The linearity of the Lagrangian in the nuclear acceleration forbids the presence of derivatives of the position of order higher than the second in the conserved energy. The complete expression is 
 {\small
\begin{align}
    &E^{\mathrm{PAW}}_{\rm R,\dot{R}}=\sum_s\frac{M_s|\dot{\mathbf{R}}_s(t)|^2}{2}+\sum_{I=1}^{N_{\rm el}}\Bigg\{\braket{\tilde{\psi}_I(t)|\hat{H}^{\rm PAW}_{\mathrm{R,\dot{R}}}|\tilde{\psi}_I(t)}\nonumber\\
    &-\sum_s\Bigg[\braket{\tilde{\psi}_I(t)|\frac{\partial \hat{H}^{\rm PAW}_{\mathrm{R,\dot{R}}}}{\partial \dot{\mathbf{R}}_s}\cdot \dot{\mathbf{R}}_s(t)+\frac{\partial \hat{H}^{\rm PAW}_{\mathrm{R,\dot{R}}}}{\partial \ddot{\mathbf{R}}_s}\cdot \ddot{\mathbf{R}}_s(t)|\tilde{\psi}_I(t)} \nonumber\\
 &-\frac{d}{dt}\left( \braket{\tilde{\psi}_I(t)|\frac{\partial \hat{H}^{\rm PAW}_{\mathrm{R,\dot{R}}}}{\partial \ddot{\mathbf{R}}_s}|\tilde{\psi}_I(t)}\right)\cdot \dot{\mathbf{R}}_s(t)+\label{eq:energy_VI}\\
 &\frac{i\hbar}{2}\Bigg(\frac{d\bra{\tilde{\psi}_I(t)}}{dt}\frac{\partial S_{\rm R, \dot{R}}}{\partial \dot{\mathbf{R}}_s}\ket{\tilde{\psi}_I(t)}-\nonumber\\
 &\bra{\tilde{\psi}_I(t)}\frac{\partial S_{\rm R, \dot{R}}}{\partial \dot{\mathbf{R}}_s}| \frac{d\ket{\tilde{\psi}_I(t)}}{dt}\Bigg)\cdot \dot{\mathbf{R}}_s(t)\Bigg]\Bigg\}.
        \nonumber
\end{align}
}
The nuclear velocity and acceleration derivatives are reported explicitly in Eqs. \eqref{eq:PAW_derivative_dotR}, \eqref{eq:PAW_derivative_ddotR} and \eqref{eq:S_derivative_dotR}. For a direct comparison between the main results, including the conserved energy, with nuclear velocity-including atomic orbitals and with the rigidly translated atomic orbitals see Table \ref{tab:hamiltonians_dynamics}.

\subsubsection*{Norm-conserving case}
In the case of a norm-conserving pseudo-potential, $S=1$ and the augmentation charges are zero $Q^s_{ij}=0$. Moreover,  for a smooth operator within the augmentation region, such as the position, the matrix elements with the all-electron and pseudo-partial waves almost coincide
\begin{equation}
\braket{\phi^{\mathbf{R}_s}_{si}|\hat{\mathbf{r}}|\phi^{\mathbf{R}_s}_{sj}}\approx \braket{\tilde{\phi}^{\mathbf{R}_s}_{si}|\hat{\mathbf{r}}|\tilde{\phi}^{\mathbf{R}_s}_{sj}}.
\label{eq:position_pseudo_all_electron}
\end{equation}
As a consequence, the Hamiltonian reduces to the one given by Ref. \cite{PhysRevLett.136.196401}, \begin{equation}
\begin{aligned}
    \hat{H}^{\mathrm{PAW-NC}}_{\mathrm{R},\mathrm{\dot{R}}}=  &\frac{\hat{\mathbf{p}}^2}{2m}+V^{\mathrm{loc}}\\&+\sum_{s}e^{\frac{i}{\hbar}m\dot{\mathbf{R}}_s(t)\cdot \hat{\mathbf{r}}}v_{s}^{\mathrm{nl}}e^{-\frac{i}{\hbar}m\dot{\mathbf{R}}_s(t)\cdot \hat{\mathbf{r}}}.
\end{aligned}
 \label{eq:PAW_Hamiltonian-NC}
\end{equation} 
The evolution of the electronic system is determined by 
\begin{equation}
    i\hbar \frac{d\ket{\tilde{\psi}_I(t)}}{dt}=\hat{H}^{\mathrm{PAW-NC}}_{\mathrm{R},\mathrm{\dot{R}}} \ket{\tilde{\psi}_I(t)},
    \label{eq:Schrodinger_PAW_NC}
\end{equation}
which guarantees Galilean invariance since the motion at constant velocity of the entire system does not cause any spurious electronic transition. Indeed, there are no terms in the Hamiltonian mixing different electronic orbitals, overcoming the paradox of the non-adiabatic Ehrenfest dynamics without the nuclear velocity-dependent phases.

In the nuclear dynamics, the nuclear acceleration contribution to the force is approximately zero $\mathbf{F}_s^{\mathrm{HF-\ddot{R}}}\approx 0$, whereas the other two reduce to
\begin{align}
&\mathbf{F}_s^{\mathrm{HF}}=-\sum_{I=1}^{N_{\rm el}}\braket{\tilde{\psi}_I(t)|\frac{\partial \hat{H}^{\mathrm{PAW-NC}}_{\mathrm{R,\dot{R}}}}{\partial \mathbf{R}_s}|\tilde{\psi}_I(t)},\label{eq:HF_force_pseudo_vel_NC_1}\\
&\mathbf{F}_s^{\mathrm{HF-\dot{R}}}=\sum_{I=1}^{N_{\rm el}}\frac{d}{dt}\braket{\tilde{\psi}_I(t)|\frac{\partial \hat{H}^{\rm PAW-NC}_{\mathrm{R},\dot{\mathrm{R}}}}{\partial \dot{\mathbf{R}}_s}|\tilde{\psi}_I(t)} \label{eq:HF_force_pseudo_vel_NC_2}.
\end{align}

These expressions give the force-constant matrix and the Born effective charges presented in Ref. \cite{PhysRevLett.136.196401}, which neglects the nuclear-acceleration contribution. 
The conserved energy is 
\begin{equation}
    \begin{aligned}
       & E^{\mathrm{PAW-NC}}_{\mathrm{R},\dot{\mathrm{R}}}=\sum_s \frac{M_s|\dot{\mathbf{R}}_s(t)|^2}{2}+
        \sum_{I=1}^{N_{\rm el}}\bra{\tilde{\psi}_I(t)}\\&\left(\hat{H}^{\rm PAW-\mathrm{NC}}_{\mathrm{R},\dot{\mathrm{R}}}-\sum_s\frac{\partial \hat{H}^{\rm PAW-NC}_{\mathrm{R},\dot{\mathrm{R}}}}{\partial \dot{\mathbf{R}}_s}\cdot \dot{\mathbf{R}}_s\right)\ket{\tilde{\psi}_I(t)}\Bigg],
    \end{aligned}
    \label{eq:energy_VI_NC}
\end{equation}
where, from Eq. \eqref{eq:PAW_derivative_dotR},
\begin{equation}
    \frac{\partial \hat{H}^{\mathrm{PAW-NC}}_{\mathrm{R},\mathrm{\dot{R}}}}{\partial \dot{\mathbf{R}}_{s}}=\frac{im}{\hbar}\left[\mathbf{r},v_s^{\mathrm{nl}}\right].
\end{equation}
that coincides with the results of Ref. \cite{PhysRevLett.136.196401}.
\section{Conclusions}\label{sec:conclusion}
In this work, we constructed the pseudo-potential Hamiltonian through a PAW transformation that employs velocity including atomic orbitals, in both the norm-conserving and ultrasoft cases, with the aim of obtaining the non-adiabatic Ehrenfest dynamics, as summarised in Table \ref{tab:hamiltonians_dynamics}. While the local part of the Hamiltonian remains unaffected, Peierls-like phases that depend on the nuclear velocity appear in the non-local part of the potential. Additional nuclear velocity and acceleration contributions, depending on the augmentation charge, are also present in the ultrasoft pseudopotential case. Indeed, the nuclear acceleration contributions, previously overlooked in the literature, cannot be neglected in the ultrasoft case, while in the norm-conserving case they are small and can be disregarded. Thus, the use of nuclear-velocity including atomic orbitals in the PAW transformation enables the proper reproduction of the all -electron properties in the pseudo-potential Hamiltonian when nuclei are moving, at any order in the nuclear velocity and addressing specific issues that arise in standard approaches. First of all, since electronic wavefunctions acquire spherical harmonics components at any order when nuclei are moving, neglecting the nuclear-velocity dependent phases in the atomic orbitals implies that the static orbital basis set is incomplete, making it inadequate for the description of the all-electron problem. Furthermore, our equations fully restore Galilean invariance through nuclear-velocity-dependent phases that allow for the removal of spurious couplings between different electronic states, which would otherwise appear also when the entire system undergoes a global translation. This is highlighted in Table \ref{tab:hamiltonians_dynamics} in the comparison between the electronic dynamics obtained with the rigid translation of electronic orbitals to the time-dependent nuclear positions and the use of the nuclear-velocity dependent orbitals. Finally, we generalise to ultrasoft pseudo-potentials the results of Ref. \cite{PhysRevLett.136.196401}, which is limited to the norm-conserving case, showing that additional nuclear velocity and acceleration dependent contributions have to be accounted for in the assessment of dynamical vibrational responses as the force constant matrix and the Born effective charge tensor.
\section*{Acknowledgements}
We acknowledge the MORE-TEM ERC-SYN project, Grant Agreement No. 951215. PF acknowledge also the funding from the project Ateneo 2025 by Sapienza - University of Rome (code: B83C25004300005). 
We thank Massimiliano Stengel, Raffaele Resta, Antimo Marrazzo and Giorgio Sangiovanni for useful discussions and suggestions.

\clearpage

\captionsetup{width=\textwidth}
\begin{table}[h!]
\centering
\begin{tabular}{c|c}
 \multicolumn{2}{c}{Static Hamiltonian}\rule{0pt}{0.25cm}\\
 \hline
 \multicolumn{2}{c}{$\begin{aligned}
    \hat{H}^{\rm PAW}=&\frac{\mathbf{p}^2}{2m}+V^{\mathrm{loc}}+\sum_{s}v_{s}^{\mathrm{nl}}
\end{aligned}$,\qquad Eq. \eqref{eq:PAW_Hamiltoniana_standard}}\rule{0pt}{0.5cm}\\
\midrule
\midrule
\rule[-0.5cm]{0pt}{1cm}  Refs. \cite{PhysRevB.73.035408, 10.1063/1.3700800, 10.1063/5.0252559} & \makecell[c]{This work \\
    with velocity-including orbitals} \\  
    \midrule
 \multicolumn{2}{c}{Norm-conserving Hamiltonian}\rule{0pt}{0.25cm}\\
\hline
\rule{0pt}{0.75cm} 
\makecell[c]{$\displaystyle
\begin{aligned}
    \hat{H}^{\mathrm{PAW-NC}}_{\rm R}=&\hat{H}^{\rm PAW}-\sum_{s,ij}\dot{\mathbf{R}}_s(t) \cdot \ket{\tilde{p}^{\mathbf{R}_s(t)}_{si}} \Delta \hat{\mathbf{p}}^s_{ij}\bra{\tilde{p}^{\mathbf{R}_s(t)}_{sj}}
\end{aligned}$,\\
\qquad Eq.\eqref{eq:paw_hamiltonian_R_NC}} & \makecell[c]{$\displaystyle \hat{H}^{\mathrm{PAW-NC}}_{\rm R,\mathrm{\dot{R}}}=  \frac{\hat{\mathbf{p}}^2}{2m}+V^{\mathrm{loc}}+\sum_{s}e^{\frac{i}{\hbar}m\dot{\mathbf{R}}_s(t)\cdot \hat{\mathbf{r}}}v_{s}^{\mathrm{nl}}e^{-\frac{i}{\hbar}m\dot{\mathbf{R}}_s(t)\cdot \hat{\mathbf{r}}}$,\\ Eq. \eqref{eq:PAW_Hamiltonian-NC} and Ref. \cite{PhysRevLett.136.196401}}\\
 \hline
 \multicolumn{2}{c}{Ultrasoft Hamiltonian}\rule[-0.1cm]{0pt}{0.5cm}\\
 \midrule
 \makecell[c]{
$\begin{aligned}
\hat{H}^{\mathrm{PAW}}_{\rm R}=\hat{H}^{\mathrm{PAW-NC}}_{\rm R}-\sum_{ij}\left\{\frac{\hat{\mathbf{p}}}{2},\ket{\tilde{p}^{\mathbf{R}_s(t)}_{si}}Q^{s}_{ij}\bra{\tilde{p}^{\mathbf{R}_s(t)}_{sj}} \right\}\end{aligned}$,\\
\\
Eqs. \eqref{eq:paw_hamiltonian_R}, \eqref{eq:M_no_vel}
 }
 & \makecell[c]{$\begin{aligned}
    & \hat{H}^{\mathrm{PAW}}_{\mathrm{R,\dot{R}}}=\hat{H}^{\mathrm{PAW-NC}}_{\rm R,\mathrm{\dot{R}}}+\\
     &+\sum_{si}\Bigg( m\ddot{\mathbf{R}}_s(t) \cdot e^{i\alpha_s(\hat{\mathbf{r}})} \ket{p^{\mathbf{R}_s(t)}_{si}}\mathbf{d}^s_{ij}\bra{p^{\mathbf{R}_s(t)}_{sj}}e^{-i\alpha_s(\hat{\mathbf{r}})} \\&
    -\frac{1}{2}\Bigg\{m \ddot{\mathbf{R}}_s(t)\cdot\left[\hat{\mathbf{r}}-\mathbf{R}_s(t)\right]-\dot{\mathbf{R}}_s(t)\cdot \hat{\mathbf{p}}+\frac{m|\dot{\mathbf{R}}_s(t)|^2}{2},\\
    &\qquad \quad e^{i\alpha_s(\hat{\mathbf{r}})} \ket{p^{\mathbf{R}_s(t)}_{si}}Q^s_{ij}\bra{p^{\mathbf{R}_s(t)}_{sj}}e^{-i\alpha_s(\hat{\mathbf{r}})} \Bigg\}\Bigg), 
 \end{aligned}$\\ Eq. \eqref{eq:PAW_Hamiltonian-Rdot}}\\
 \midrule
 \multicolumn{2}{c}{Schrödinger equation for the non-adiabatic Ehrenfest dynamics for ultrasoft pseudopotentials}\rule[-0.25cm]{0pt}{0.5cm}\\
\hline
\hline 
 \rule[-1.2cm]{0pt}{2.4cm}   
 \makecell[c]{$   \displaystyle \begin{aligned}
 &\displaystyle i\hbar S \frac{d\ket{\tilde{\psi}_I(t)}}{dt}=\left(\hat{H}^{\mathrm{PAW}}_{\rm R}-\frac{1}{2}\frac{dS}{dt}\right) \ket{\tilde{\psi}_I(t)}
\end{aligned}
$, \qquad 
Eq. \eqref{eq:Schrodinger_PAW_no_vel}\\
where \\
$\begin{aligned}S=&\mathbb{1}+\sum_{s,ij}\ket{\tilde{p}^{\mathbf{R}_s(t)}_{si}}Q^s_{ij}\bra{\tilde{p}^{\mathbf{R}_s(t)}_{sj}}\end{aligned}$, \qquad Eq. \eqref{eq:TdaggerT_identity}}
&\makecell[c]{$\displaystyle  i\hbar S_{\rm R,\mathrm{\dot{R}}}\frac{d\ket{\tilde{\psi}_I(t)}}{dt}=\left(\hat{H}^{\mathrm{PAW}}_{\mathrm{R,\dot{R}}}-\frac{1}{2}\frac{dS_{\rm R,\mathrm{\dot{R}}}}{dt}\right)\ket{\tilde{\psi}_I(t)}$, \qquad Eq. \eqref{eq:Schrodinger_PAW}\\ 
 where \\
$S_{\rm R,\mathrm{\dot{R}}}=\mathbb{1}+\sum_{s,ij}e^{i\alpha_s(\hat{\mathbf{r}})}\ket{\tilde{p}^{\mathbf{R}_s(t)}_{si}}Q^s_{ij}\bra{\tilde{p}^{\mathbf{R}_s(t)}_{sj}}e^{-i\alpha_s(\hat{\mathbf{r}})}$,  Eq. \eqref{eq:TdaggerT_identity_vel}
 }\\
 \multicolumn{2}{c}{\makecell[c]{ with \qquad \\
 $Q^s_{ij}=\braket{\phi^{\mathbf{R}_s}_{si}|\phi^{\mathbf{R}_s}_{sj}}-\braket{\tilde{\phi}^{\mathbf{R}_s}_{si}|\tilde{\phi}^{\mathbf{R}_s}_{sj}}$\\
 \newline
 In the norm-conserving case $\displaystyle S=\mathbb{1}$,$\displaystyle \frac{dS}{dt}=0$ (see Eqs. \eqref{eq:Schrodinger_PAW_no_vel_NC}, \eqref{eq:Schrodinger_PAW_NC})}}\rule[-1cm]{0pt}{2cm}  \\

\midrule
\midrule

\multicolumn{2}{c}{Conserved energy}\rule{0pt}{0.35cm}\\
\midrule
\makecell[c]{$\begin{aligned}
E^{\mathrm{PAW}}_{\rm R}=\sum_s\frac{M_s}{2}|\dot{\mathbf{R}}_s(t)|^2+\sum_{I=1}^{N_{\rm el}}\braket{\tilde{\psi}_I(t)|\hat{H}^{\rm PAW}|\tilde{\psi}_I(t)}
    \end{aligned}$,\\ \\ Eq. \eqref{eq:conserved_energy_PAW}}& 
    \makecell[c]{$\begin{aligned}
    &E^{\mathrm{PAW}}_{\rm R,\dot{R}}=\sum_s\frac{M_s|\dot{\mathbf{R}}_s(t)|^2}{2}+\sum_{I=1}^{N_{\rm el}}\Bigg\{\braket{\tilde{\psi}_I(t)|\hat{H}^{\rm PAW}_{\mathrm{R,\dot{R}}}|\tilde{\psi}_I(t)}\\
 &-\sum_s\Bigg[\braket{\tilde{\psi}_I(t)|\frac{\partial \hat{H}^{\rm PAW}_{\mathrm{R,\dot{R}}}}{\partial \dot{\mathbf{R}}_s}\cdot \dot{\mathbf{R}}_s(t)+\frac{\partial \hat{H}^{\rm PAW}_{\mathrm{R,\dot{R}}}}{\partial \ddot{\mathbf{R}}_s}\cdot \ddot{\mathbf{R}}_s(t)|\tilde{\psi}_I(t)} \\
 &-\frac{d}{dt}\left( \braket{\tilde{\psi}_I(t)|\frac{\partial \hat{H}^{\rm PAW}_{\mathrm{R,\dot{R}}}}{\partial \ddot{\mathbf{R}}_s}|\tilde{\psi}_I(t)}\right)\cdot \dot{\mathbf{R}}_s(t)+\frac{i\hbar}{2}\Bigg(\frac{d\bra{\tilde{\psi}_I(t)}}{dt}\frac{\partial S_{\rm R, \dot{R}}}{\partial \dot{\mathbf{R}}_s}\ket{\tilde{\psi}_I(t)}\\&-\bra{\tilde{\psi}_I(t)}\frac{\partial S_{\rm R, \dot{R}}}{\partial \dot{\mathbf{R}}_s}| \frac{d\ket{\tilde{\psi}_I(t)}}{dt}\Bigg)\cdot \dot{\mathbf{R}}_s(t)\Bigg]\Bigg\},
        \end{aligned}$\\ Eq. \eqref{eq:energy_VI} and Ref. \cite{PhysRevLett.136.196401} for the norm-conserving limit}
\end{tabular}
\caption{Comparison between the Hamiltonians in the ultrasoft and norm-conserving case; the Schrödinger equation for the electronic non-adiabatic Ehrenfest dynamics; the conserved energy obtained either rigidly shifting the electronic orbitals with the nuclei (left column and Refs. \cite{PhysRevB.73.035408, 10.1063/1.3700800, 10.1063/5.0252559}) or applying the nuclear velocity dependent phases to the electronic wavefunction (right column).}
\label{tab:hamiltonians_dynamics}
    \end{table}

\clearpage

\appendix
\begin{widetext}
\section{Temporal evolution with a non-Hermitian Hamiltonian and norm-conservation}
\label{app:temporal}
Consider a system where the electronic part of the Lagrangian is
\begin{equation}
    \begin{aligned}
    &\mathcal{L}_{\rm el}=-\sum_{I=1}^{N_{\rm el}}\Bigg[\braket{\tilde{\psi}_I(t)|\hat{O}(t)|\tilde{\psi}_I(t)}+\frac{i\hbar}{2}\left(\frac{d\bra{\tilde{\psi}_I(t)}}{dt}S\ket{\tilde{\psi}_I(t)}-\bra{\tilde{\psi}_I(t)}S| \frac{d\ket{\tilde{\psi}_I(t)}}{dt}\right)\Bigg]
\end{aligned}
\end{equation}
and the the wavefunction is normalized such that 
\begin{equation}
    \braket{\psi(t)|S|\psi(t)}=1,
\end{equation}
then the temporal evolution operator $\hat{O}(t)$ is non-hermitian operator 
\begin{equation}
    i\hbar S {\ket{\dot{\psi}(t)}}=O(t)\ket{\psi(t)}.
\end{equation}

In order to prove this, we compute the temporal derivative of the norm
\begin{equation}
\begin{aligned}
      &\frac{d}{dt}\left(\braket{\psi(t)|S|\psi(t)}\right)=0,\\
      &\braket{\dot{\psi}(t)|S|\psi(t)}+\braket{\psi(t)|S|\dot{\psi}(t)}+\braket{\psi(t)|\frac{dS(t)}{dt}|\psi(t)}=0.\\
\end{aligned}
\end{equation}
Substituting the Schrödinger equations for $\ket{\psi}$ and $\bra{\psi}$, we obtain
\begin{equation}
\begin{aligned}
      \braket{\psi|-O(t)+O^{\dagger}(t)+\frac{dS(t)}{dt}|\psi}=0,\\
\end{aligned}
\end{equation}
following that 
\begin{equation}
    \frac{dS(t)}{dt}=O(t)-O^{\dagger}(t). 
\end{equation}
In conclusion, a time-dependent $S$ implies that the evolution is not determined by a Hermitian operator.  We remark that the evolution is unitary because of the definition of the norm. A non-Hermitian evolution operator is needed to conserve the norm. 
\section{Details on the transformation of operators with the velocity-including PAW}
\label{app:details_operators}
According to Eq. \eqref{eq:otildevipaw} of the main text, here reported, the operators transforms with the velocity-including PAW transformation as 
\begin{equation}
\begin{aligned}
      &\tilde{O}=O+\sum_{s,ij} e^{i\alpha_s(\hat{\mathbf{r}})}\ket{\tilde{p}^{\mathbf{R}_s(t)}_{si}} \Delta O^{s, \mathrm{\dot{R}}}_{ij}\bra{\tilde{p}^{\mathbf{R}_s(t)}_{sj}}e^{-i\alpha_s(\hat{\mathbf{r}})},\\
     &\Delta O^{s, \mathrm{\dot{R}}}_{ij}=\braket{\phi^{\mathbf{R}_s(t)}_{si}|e^{-i\alpha_s(\hat{\mathbf{r}})}Oe^{i\alpha_s(\hat{\mathbf{r}})}|\phi^{\mathbf{R}_s(t)}_{sj}}
     -\braket{\tilde{\phi}^{\mathbf{R}_s(t)}_{si}|e^{-i\alpha_s(\hat{\mathbf{r}})}Oe^{i\alpha_s(\hat{\mathbf{r}})}|\tilde{\phi}^{\mathbf{R}_s(t)}_{sj}}.
     \label{eq:otildevipaw_app}
\end{aligned}
\end{equation}
We remark that $\Delta O^{s, \mathrm{\dot{R}}}_{ij}$ and $\Delta O^{s}_{ij}$ of Eq. \eqref{eq:delta_O} of the main text indicate the difference between all-electron and pseudo wavefunctions matrix elements with and without the nuclear velocity dependent phases in the orbitals, respectively.
Adding and subtracting $\displaystyle \sum_{s,ij} e^{i\alpha_s(\hat{\mathbf{r}})}\ket{\tilde{p}^{\mathbf{R}_s(t)}_{si}} \Delta O^{s}_{ij}\bra{\tilde{p}^{\mathbf{R}_s(t)}_{sj}}e^{-i\alpha_s(\hat{\mathbf{r}})}$, we can rewrite Eq. \eqref{eq:otildevipaw_app} as
 \begin{equation}
     \begin{aligned}
           &\tilde{O}=O+\sum_{s,ij} \Bigg(e^{i\alpha_s(\hat{\mathbf{r}})}\ket{\tilde{p}^{\mathbf{R}_s(t)}_{si}} \Delta O^{s}_{ij}\bra{\tilde{p}^{\mathbf{R}_s(t)}_{sj}}e^{-i\alpha_s(\hat{\mathbf{r}})}+e^{i\alpha_s(\hat{\mathbf{r}})}\ket{\tilde{p}^{\mathbf{R}_s(t)}_{si}}\left( \Delta O^{s,\mathrm{\dot{R}}}_{ij}-\Delta O^{s}_{ij}\right)\bra{\tilde{p}^{\mathbf{R}_s(t)}_{sj}}e^{-i\alpha_s(\hat{\mathbf{r}})}\Bigg).
     \end{aligned}
     \label{eq:delta_O_dotR}
 \end{equation}
where $\Delta O^{s}_{ij}$ is the difference between the pseudo and all-electron matrix elements without the nuclear velocity-dependent phases, as defined in Eq. \eqref{eq:delta_O}. The last term contains the difference between the velocity-including and the standard cases. It is expressed as 
\begin{equation}
    \begin{aligned}
\Delta O^{s,\mathrm{\dot{R}}}_{ij}-        \Delta O^s_{ij}=\braket{\phi^{\mathbf{R}_s(t)}_{si}|e^{-i\alpha_s(\hat{\mathbf{r}})}Oe^{i\alpha_s(\hat{\mathbf{r}})}-O|\phi^{\mathbf{R}_s(t)}_{sj}}-\braket{\tilde{\phi}^{\mathbf{R}_s(t)}_{si}|e^{-i\alpha_s(\hat{\mathbf{r}})}Oe^{i\alpha_s(\hat{\mathbf{r}})}-O|\tilde{\phi}^{\mathbf{R}_s(t)}_{sj}}
    \end{aligned}
\end{equation}
For instance, for an operator depending on the momentum $O(\hat{\mathbf{p}})$, the nuclear velocity-dependent phase factor cause a shift in momentum, according to the commutation rules between position and momentum operators $e^{-i\alpha_s(\hat{\mathbf{r}})}O(\hat{\mathbf{p}})e^{is\alpha_s(\hat{\mathbf{r}})}=O(\hat{\mathbf{p}}+m\dot{\mathbf{R}}_s(t))$, implying that
\begin{equation}
    \begin{aligned}
\Delta O^{s,\mathrm{\dot{R}}}_{ij}-        \Delta O^s_{ij}=\braket{\phi^{\mathbf{R}_s(t)}_{si}|O(\hat{\mathbf{p}}+m\dot{\mathbf{R}}_s(t))-O|\phi^{\mathbf{R}_s(t)}_{sj}}-\braket{\tilde{\phi}^{\mathbf{R}_s(t)}_{si}|O(\hat{\mathbf{p}}+m\dot{\mathbf{R}}_s(t))-O|\tilde{\phi}^{\mathbf{R}_s(t)}_{sj}}.
    \end{aligned}
\end{equation}

\section{Detailed calculation of the temporal derivative of the PAW transformation}\label{app:computation}
The PAW transformation of Eq. \eqref{eq:PAW_transformation} for moving nuclei, but without the nuclear velocity-dependent phase factors, can be expressed as 
\begin{equation}
\begin{aligned}
\hat{\mathcal{T}}=\hat{\mathbb{1}}+\sum_{s}\hat{t}^s,\qquad
\hat{t}^s=\sum_i\left(\ket{\phi^{\mathbf{R}_s(t)}_{si}}- \ket{\tilde{\phi}^{\mathbf{R}_s(t)}_{si}}\right)\bra{\tilde{p}^{\mathbf{R}_s(t)}_{si}},
\end{aligned}
\end{equation}
whereas with the nuclear velocity-dependent phases, it is expressed as in Eq. \eqref{eq:Trdot}
\begin{equation}
\begin{aligned}
     &\hat{\mathcal{T}}_{\dot{\mathrm{R}}}=\hat{\mathbb{1}}+\sum_{s}\hat{t}^s_{\dot{\mathbf{R}}_s},\qquad \hat{t}^s_{\dot{\mathbf{R}}_s}=\sum_ie^{i\alpha_s(\hat{\mathbf{r}})} 
    \left(\ket{\phi^{\mathbf{R}_s(t)}_{si}}- \ket{\tilde{\phi}^{\mathbf{R}_s(t)}_{si}}\right)\bra{\tilde{p}^{\mathbf{R}_s(t)}_{si}}e^{-i\alpha_s(\hat{\mathbf{r}})}.
     \label{eq:Trdot_app}
\end{aligned}
\end{equation}
By using the equations above, we can also express the difference between the pseudo and the all-electron (local o semi-local) operators as 
\begin{equation}
    \tilde{O}-O=\sum_s \left[(\hat{t}_{\dot{\mathbf{R}}_s}^s)^{\dagger} O+O\hat{t}_{\dot{\mathbf{R}}_s}^s+(\hat{t}_{\dot{\mathbf{R}}_s}^s)^{\dagger}O\hat{t}_{\dot{\mathbf{R}}_s}^s\right].
    \label{eq:transformation_detail}
\end{equation}
By applying the identity for the ket and the bra of the pseudo-wavefunction given in Eq. \eqref{eq:identity_augmentation_region} to the first and the second term of the right-hand side, respectively, the expression for the transformation of the PAW operators, given in Eq. \eqref{eq:PAW_operator_transformation}, is obtained. 

In the calculation of the temporal derivatives of the transformation, we consider the nuclear velocity-including case since the other case is recovered by setting $\alpha(\hat{\mathbf{r}})=0$.  
Therefore, using Eq. \eqref{eq:derivative_vi_atomic_orbital_v2} and defining 
\begin{equation}
    \mathbf{D}_s(\hat{\mathbf{r}},\hat{\mathbf{p}})=\frac{i}{\hbar}\left(m\mathbf{\ddot{R}}_s(t)\cdot (\hat{\mathbf{r}}-\mathbf{R}_s(t))-\dot{\mathbf{R}}_s(t)\cdot \mathbf{\hat{p}}\right),
    \label{eq:orbital_derivative_app}
\end{equation}
we obtain that
\begin{equation}
    \frac{d\hat{\mathcal{T}}_{\dot{\mathrm{R}}}}{dt}=  \sum_s  \left[\mathbf{D}_s(\hat{\mathbf{r}},\hat{\mathbf{p}}),\hat{t}^s_{\dot{\mathbf{R}}_s}\right].
    \label{eq:derivative_T_velocity_including_dotR_app}
\end{equation}
In the absence of the nuclear velocity-dependent phase factors, we have the same expression with $\mathbf{\ddot{R}}_s(t)=0$. 
From Eq. \eqref{eq:derivative_T_velocity_including_dotR_app}, it follows that 
\begin{equation}
    \begin{aligned}
        &\hat{\mathcal{T}}_{\dot{\mathrm{R}}}^{\dagger}\frac{d\hat{\mathcal{T}}_{\dot{\mathrm{R}}}}{dt}=\hat{\mathcal{T}}_{\dot{\mathrm{R}}}^{\dagger}\sum_s  \left[\mathbf{D}_s(\hat{\mathbf{r}},\hat{\mathbf{p}}),\hat{t}^s_{\dot{\mathbf{R}}_s}\right]
=\sum_s \left(\mathbf{D}_s\hat{t}^s_{\dot{\mathbf{R}}_s}-\hat{t}^s_{\dot{\mathbf{R}}_s}\mathbf{D}_s+  (\hat{t}^s_{\dot{\mathbf{R}}_s})^{\dagger}\mathbf{D}_s\hat{t}^s_{\dot{\mathbf{R}}_s}-(\hat{t}^s_{\dot{\mathbf{R}}_s})^{\dagger}\hat{t}^s_{\dot{\mathbf{R}}_s}\mathbf{D}_s\right).
    \end{aligned}
\end{equation}
By summing and subtracting $\displaystyle \sum_s (\hat{t}^s_{\dot{\mathbf{R}}_s})^{\dagger}\mathbf{D}_s$ 

\begin{equation}
    \begin{aligned}
&\hat{\mathcal{T}}_{\dot{\mathrm{R}}}^{\dagger}\frac{d\hat{\mathcal{T}}_{\dot{\mathrm{R}}}}{dt}
=\sum_s \bigg[\left((\hat{t}^s_{\dot{\mathbf{R}}_s})^{\dagger}\mathbf{D}_s+\mathbf{D}_s\hat{t}^s_{\dot{\mathbf{R}}_s}+  (\hat{t}^s_{\dot{\mathbf{R}}_s})^{\dagger}\mathbf{D}_s\hat{t}^s_{\dot{\mathbf{R}}_s} \right)-\left((\hat{t}^s_{\dot{\mathbf{R}}_s})^{\dagger}+\hat{t}^s_{\dot{\mathbf{R}}_s}+ (\hat{t}^s_{\dot{\mathbf{R}}_s})^{\dagger}\hat{t}^s_{\dot{\mathbf{R}}_s}\right)\mathbf{D}_s\bigg].
\end{aligned}
\end{equation}
According to Eq. \eqref{eq:transformation_detail}, the two terms of the right-hand side correspond to the augmentation region resolved difference between the all-electron and pseudo- operators $\mathbf{D}_s$ and $\mathbb{1}$. Therefore, by properly applying the identities in the augmentation region given in Eq. \eqref{eq:identity_augmentation_region}, as for the transformation of the operators, we obtain with the hypothesis that augmentations regions do not overlap and performing the same calculation for $\displaystyle \frac{d\hat{\mathcal{T}}^{\dagger}_{\mathrm{\dot{R}}}}{dt}\hat{\mathcal{T}}_{\dot{\mathrm{R}}}$,
\begin{equation}
    \begin{aligned}
        \hat{\mathcal{T}}_{\dot{\mathrm{R}}}^{\dagger}\frac{d\hat{\mathcal{T}}_{\dot{\mathrm{R}}}}{dt}= &\sum_{sij} e^{i\alpha_s(\hat{\mathbf{r}})} \Big(\ket{p^{\mathbf{R}_s(t)}_{si}}\Delta \mathbf{D}^{s,\mathrm{\dot{R}}}_{ij}\bra{p^{\mathbf{R}_s(t)}_{sj}}e^{-i\alpha_s(\hat{\mathbf{r}})}  -\ket{p^{\mathbf{R}_s(t)}_{si}}Q^s_{ij}\bra{p^{\mathbf{R}_s(t)}_{sj}}e^{-i\alpha_s(\hat{\mathbf{r}})} \mathbf{D}_s\Big)\\
        \frac{d\hat{\mathcal{T}}^{\dagger}_{\mathrm{\dot{R}}}}{dt}\hat{\mathcal{T}}_{\dot{\mathrm{R}}}=& \sum_{sij} \Big(-e^{i\alpha_s(\hat{\mathbf{r}})} \ket{p^{\mathbf{R}_s(t)}_{si}}\Delta \mathbf{D}^{s,\mathrm{\dot{R}}}_{ij}\bra{p^{\mathbf{R}_s(t)}_{sj}}+\mathbf{D}_se^{i\alpha_s(\hat{\mathbf{r}})} \ket{p^{\mathbf{R}_s(t)}_{si}}Q^s_{ij}\bra{p^{\mathbf{R}_s(t)}_{sj}}\Big)e^{i\alpha_s(\hat{\mathbf{r}})}.
    \end{aligned}
    \label{eq:TdotT_app}
\end{equation}
where, by using that $e^{-i\alpha_{s}(\hat{\mathbf{r}})}\hat{\mathbf{p}}e^{i\alpha_{s}(\hat{\mathbf{r}})}=\hat{\mathbf{p}}+m\dot{\mathbf{R}}_s(t)$ in combination with the definition of $\mathbf{D}_s$ given in  Eq. \eqref{eq:orbital_derivative_app},
\begin{equation}
\begin{aligned}
    \Delta \mathbf{D}^{s,\mathrm{\dot{R}}}_{ij}=&\frac{i}{\hbar}\Bigg(\braket{\phi^{\mathbf{R}_s(t)}_{si}|m\mathbf{\ddot{R}}_s(t)\cdot (\hat{\mathbf{r}}-\mathbf{R}_s(t))-\dot{\mathbf{R}}_s(t)\cdot \mathbf{\hat{p}}-m|\dot{\mathbf{R}}_s(t)|^2|\phi^{\mathbf{R}_s(t)}_{sj}}
\\&-\braket{\tilde{\phi}^{\mathbf{R}_s(t)}_{si}|m\mathbf{\ddot{R}}_s(t)\cdot (\hat{\mathbf{r}}-\mathbf{R}_s(t))-\dot{\mathbf{R}}_s(t)\cdot \mathbf{\hat{p}}-m|\dot{\mathbf{R}}_s(t)|^2|\tilde{\phi}^{\mathbf{R}_s(t)}_{sj}}\Bigg).
\end{aligned}
\end{equation}

Finally, the contribution to the Lagrangian originating from the temporal derivative of the transformation is
\begin{equation}
\begin{aligned}
&\frac{i\hbar}{2}\left(  \frac{d\hat{\mathcal{T}}_{\dot{\mathrm{R}}}^{\dagger}}{dt}\hat{\mathcal{T}}_{\dot{\mathrm{R}}}-\hat{\mathcal{T}}_{\dot{\mathrm{R}}}^{\dagger}\frac{d\hat{\mathcal{T}}_{\dot{\mathrm{R}}}}{dt}\right)= -i\hbar\sum_{s,ij} \Bigg(e^{i\alpha_s(\hat{\mathbf{r}})} \ket{\tilde{p}^{\mathbf{R}_s(t)}_{sj}} \Delta \mathbf{D}^{s,\mathrm{\dot{R}}}_{ij}\bra{\tilde{p}^{\mathbf{R}_s(t)}_{si}}
  e^{-i\alpha_s(\hat{\mathbf{r}})}-\left\{\frac{\mathbf{D}_s}{2},e^{i\alpha_s(\hat{\mathbf{r}})} \ket{p^{\mathbf{R}_s(t)}_{si}}Q^s_{ij}\bra{p^{\mathbf{R}_s(t)}_{sj}}e^{-i\alpha_s(\hat{\mathbf{r}})}\right\}\Bigg).
\end{aligned}
\label{eq:acceleration_PAW_lagrangian_app}
\end{equation} 
In the norm-conserving case, where $Q^s_{ij}=0$ and $\hat{\mathcal{T}}_{\dot{\mathrm{R}}}^{\dagger}\hat{\mathcal{T}}_{\dot{\mathrm{R}}}=\mathbb{1}$, the temporal derivative terms reduces to 
\begin{equation}
    \begin{aligned}
        \hat{\mathcal{T}}_{\dot{\mathrm{R}}}^{\dagger}\frac{d\hat{\mathcal{T}}_{\dot{\mathrm{R}}}}{dt}=-\frac{d\hat{\mathcal{T}^{\dagger}}_{\{\dot{\mathbf{R}}_s\}}}{dt}\hat{\mathcal{T}}_{\dot{\mathrm{R}}}= &\sum_s e^{i\alpha_s(\hat{\mathbf{r}})} \ket{p^{\mathbf{R}_s(t)}_{si}}\Delta \mathbf{D}^{s,\mathrm{\dot{R}}}_{ij}\bra{p^{\mathbf{R}_s(t)}_{sj}}e^{-i\alpha_s(\hat{\mathbf{r}})} 
    \end{aligned}
    \label{eq:TdotT_app_norm_conserving}
\end{equation}
In the absence of nuclear velocity-dependent phase factors, the proof is the same with $\mathbf{\ddot{R}}_s(t)=0$ and $\Delta \mathbf{D}^{s,\mathrm{\dot{R}}}_{ij}$ replaced by $\Delta \mathbf{D}^{s}_{ij}$, which gives a result Eq. \eqref{eq:M_no_vel} of the main text.

\end{widetext}

\bibliography{biblio}

@article{ostrogradsky1850memoires,
  author  = {Ostrogradsky, Mikhail},
  title   = {M{\'e}moires sur les {\'e}quations diff{\'e}rentielles, relatives au probl{\`e}me des isop{\'e}rim{\`e}tres},
  journal = {M{\'e}moires de l'Acad{\'e}mie Imp{\'e}riale des Sciences de St.-P{\'e}tersbourg, S{\'e}rie 6},
  volume  = {4},
  year    = {1850},
  pages   = {385--517}
}

@article{Bates1958_electroncapture,
  author    = {Bates, D. R. and McCarroll, R.},
  title     = {Electron Capture in Slow Collisions},
  journal   = {Proceedings of the Royal Society of London. Series A, Mathematical and Physical Sciences},
  volume    = {245},
  number    = {1241},
  pages     = {175--183},
  year      = {1958},
  publisher = {The Royal Society},
  url       = {http://www.jstor.org/stable/100394},
}

@article{PhysRevA.18.117,
  title = {Theory of near-adiabatic collisions. I. Electron translation factor method},
  author = {Thorson, W. R. and Delos, J. B.},
  journal = {Phys. Rev. A},
  volume = {18},
  issue = {1},
  pages = {117--134},
  numpages = {0},
  year = {1978},
  month = {Jul},
  publisher = {American Physical Society},
  doi = {10.1103/PhysRevA.18.117},
  url = {https://link.aps.org/doi/10.1103/PhysRevA.18.117}
}

@article{PhysRevLett.43.1494,
  title = {Norm-Conserving Pseudopotentials},
  author = {Hamann, D. R. and Schl\"uter, M. and Chiang, C.},
  journal = {Phys. Rev. Lett.},
  volume = {43},
  issue = {20},
  pages = {1494--1497},
  numpages = {0},
  year = {1979},
  month = {11},
  publisher = {American Physical Society},
  doi = {10.1103/PhysRevLett.43.1494},
  url = {https://link.aps.org/doi/10.1103/PhysRevLett.43.1494}
}

@article{PhysRevB.26.4199,
  title = {Pseudopotentials that work: From H to Pu},
  author = {Bachelet, G. B. and Hamann, D. R. and Schl\"uter, M.},
  journal = {Phys. Rev. B},
  volume = {26},
  issue = {8},
  pages = {4199--4228},
  numpages = {0},
  year = {1982},
  month = {10},
  publisher = {American Physical Society},
  doi = {10.1103/PhysRevB.26.4199},
  url = {https://link.aps.org/doi/10.1103/PhysRevB.26.4199}
}

@article{PhysRevLett.48.1425,
  title = {Efficacious Form for Model Pseudopotentials},
  author = {Kleinman, Leonard and Bylander, D. M.},
  journal = {Phys. Rev. Lett.},
  volume = {48},
  issue = {20},
  pages = {1425--1428},
  numpages = {0},
  year = {1982},
  month = {May},
  publisher = {American Physical Society},
  doi = {10.1103/PhysRevLett.48.1425},
  url = {https://link.aps.org/doi/10.1103/PhysRevLett.48.1425}
}

@article{10.1063/1.445588,
    author = {Nafie, Laurence A.},
    title = {Adiabatic molecular properties beyond the Born–Oppenheimer approximation. Complete adiabatic wave functions and vibrationally induced electronic current density},
    journal = {The Journal of Chemical Physics},
    volume = {79},
    number = {10},
    pages = {4950-4957},
    year = {1983},
    month = {11},
    issn = {0021-9606},
    doi = {10.1063/1.445588},
    url = {https://doi.org/10.1063/1.445588}
}

@article{PhysRevA.23.2301,
  title = {Theory of near-adiabatic collisions. III. Coupled equations arising from expansions involving single-center states},
  author = {Delos, J. B.},
  journal = {Phys. Rev. A},
  volume = {23},
  issue = {5},
  pages = {2301--2318},
  numpages = {0},
  year = {1981},
  month = {May},
  publisher = {American Physical Society},
  doi = {10.1103/PhysRevA.23.2301},
  url = {https://link.aps.org/doi/10.1103/PhysRevA.23.2301}
}

@article{RJAllan_1985,
doi = {10.1088/0022-3700/18/10/015},
url = {https://doi.org/10.1088/0022-3700/18/10/015},
year = {1985},
month = {may},
publisher = {},
volume = {18},
number = {10},
pages = {1981},
author = {R J Allan and J Hanssen},
title = {Quasimolecular treatment of Na-Na+, Li-Li+, Li-Na+ and Na-Li+ collisions with a common translation factor},
journal = {Journal of Physics B: Atomic and Molecular Physics}
}

@article{PhysRevB.59.1758,
  title = {From ultrasoft pseudopotentials to the projector augmented-wave method},
  author = {Kresse, G. and Joubert, D.},
  journal = {Phys. Rev. B},
  volume = {59},
  issue = {3},
  pages = {1758--1775},
  numpages = {0},
  year = {1999},
  month = {Jan},
  publisher = {American Physical Society},
  doi = {10.1103/PhysRevB.59.1758},
  url = {https://link.aps.org/doi/10.1103/PhysRevB.59.1758}
}

@article{doi:10.1021/jp9906839,
author = {Micha, David A.},
title = {Time-Dependent Many-Electron Treatment of Electronic Energy and Charge Transfer in Atomic Collisions},
journal = {The Journal of Physical Chemistry A},
volume = {103},
number = {38},
pages = {7562-7574},
year = {1999},
doi = {10.1021/jp9906839},
URL = {https://doi.org/10.1021/jp9906839}
}

@article{RJAllan1985bis,
doi = {10.1088/0022-3700/18/10/016},
url = {https://doi.org/10.1088/0022-3700/18/10/016},
year = {1985},
month = {may},
publisher = {},
volume = {18},
number = {10},
pages = {1999},
author = {R J Allan and A Bahring and J Hanssen},
title = {Alignment and orientation of atomic orbitals in Na(3p)+Na+ interactions},
journal = {Journal of Physics B: Atomic and Molecular Physics}
}

@article{PhysRevA.35.70,
  title = {Traveling-molecular-orbital-expansion studies of electron capture in collisions of fully stripped ions (Z=6--9) with H and ${\mathrm{H}}_{2}$},
  author = {Kimura, M. and Lane, N. F.},
  journal = {Phys. Rev. A},
  volume = {35},
  issue = {1},
  pages = {70--78},
  numpages = {0},
  year = {1987},
  month = {Jan},
  publisher = {American Physical Society},
  doi = {10.1103/PhysRevA.35.70},
  url = {https://link.aps.org/doi/10.1103/PhysRevA.35.70}
}

@article{PhysRevB.41.7892,
  title = {Soft self-consistent pseudopotentials in a generalized eigenvalue formalism},
  author = {Vanderbilt, David},
  journal = {Phys. Rev. B},
  volume = {41},
  issue = {11},
  pages = {7892--7895},
  numpages = {0},
  year = {1990},
  month = {Apr},
  publisher = {American Physical Society},
  doi = {10.1103/PhysRevB.41.7892},
  url = {https://link.aps.org/doi/10.1103/PhysRevB.41.7892}
}

@article{PhysRevB.43.1993,
  title = {Efficient pseudopotentials for plane-wave calculations},
  author = {Troullier, N. and Martins, Jos\'e Lu\'{\i}s},
  journal = {Phys. Rev. B},
  volume = {43},
  issue = {3},
  pages = {1993--2006},
  numpages = {0},
  year = {1991},
  month = {1},
  publisher = {American Physical Society},
  doi = {10.1103/PhysRevB.43.1993},
  url = {https://link.aps.org/doi/10.1103/PhysRevB.43.1993}
}

@article{10.1063/1.462668,
    author = {Nafie, Laurence A.},
    title = {Velocity‐gauge formalism in the theory of vibrational circular dichroism and infrared absorption},
    journal = {The Journal of Chemical Physics},
    volume = {96},
    number = {8},
    pages = {5687-5702},
    year = {1992},
    month = {04},
    issn = {0021-9606},
    doi = {10.1063/1.462668},
    url = {https://doi.org/10.1063/1.462668}
}

@article{PhysRevB.47.10142,
  title = {Car-Parrinello molecular dynamics with Vanderbilt ultrasoft pseudopotentials},
  author = {Laasonen, Kari and Pasquarello, Alfredo and Car, Roberto and Lee, Changyol and Vanderbilt, David},
  journal = {Phys. Rev. B},
  volume = {47},
  issue = {16},
  pages = {10142--10153},
  numpages = {0},
  year = {1993},
  month = {Apr},
  publisher = {American Physical Society},
  doi = {10.1103/PhysRevB.47.10142},
  url = {https://link.aps.org/doi/10.1103/PhysRevB.47.10142}
}

@article{PhysRevA.50.418,
  title = {Common-translation-factor method with an atomic basis},
  author = {Errea, L. F. and M\'endez, L. and Riera, A. and Harel, C. and Jouin, H. and Pons, B.},
  journal = {Phys. Rev. A},
  volume = {50},
  issue = {1},
  pages = {418--422},
  numpages = {0},
  year = {1994},
  month = {Jul},
  publisher = {American Physical Society},
  doi = {10.1103/PhysRevA.50.418},
  url = {https://link.aps.org/doi/10.1103/PhysRevA.50.418}
}

@article{LFErrea_1994,
doi = {10.1088/0953-4075/27/16/010},
url = {https://doi.org/10.1088/0953-4075/27/16/010},
year = {1994},
month = {aug},
publisher = {},
volume = {27},
number = {16},
pages = {3603},
author = {L F Errea and C Harel and H Jouini and L Mendez and B Pons and A Riera},
title = {Common translation factor method},
journal = {Journal of Physics B: Atomic, Molecular and Optical Physics}
}

@article{PhysRevB.50.17953,
  title = {Projector augmented-wave method},
  author = {Bl\"ochl, P. E.},
  journal = {Phys. Rev. B},
  volume = {50},
  issue = {24},
  pages = {17953--17979},
  numpages = {0},
  year = {1994},
  month = {12},
  publisher = {American Physical Society},
  doi = {10.1103/PhysRevB.50.17953},
  url = {https://link.aps.org/doi/10.1103/PhysRevB.50.17953}
}

@article{PhysRevB.56.R11369,
  title = {Density-functional perturbation theory for lattice dynamics with ultrasoft pseudopotentials},
  author = {Dal Corso, Andrea and Pasquarello, Alfredo and Baldereschi, Alfonso},
  journal = {Phys. Rev. B},
  volume = {56},
  issue = {18},
  pages = {R11369--R11372},
  numpages = {0},
  year = {1997},
  month = {Nov},
  publisher = {American Physical Society},
  doi = {10.1103/PhysRevB.56.R11369},
  url = {https://link.aps.org/doi/10.1103/PhysRevB.56.R11369}
}

@article{PhysRevLett.80.3029,
  title = {Classical Outlook on the Electron Translation Factor Problem},
  author = {Illescas, Clara and Riera, A.},
  journal = {Phys. Rev. Lett.},
  volume = {80},
  issue = {14},
  pages = {3029--3032},
  numpages = {0},
  year = {1998},
  month = {Apr},
  publisher = {American Physical Society},
  doi = {10.1103/PhysRevLett.80.3029},
  url = {https://link.aps.org/doi/10.1103/PhysRevLett.80.3029}
}

@article{PhysRevA.62.022703,
  title = {Time-dependent many-electron approach to slow ion-atom collisions for systems with several active electrons},
  author = {Runge, Keith and Micha, David A.},
  journal = {Phys. Rev. A},
  volume = {62},
  issue = {2},
  pages = {022703},
  numpages = {10},
  year = {2000},
  month = {Jul},
  publisher = {American Physical Society},
  doi = {10.1103/PhysRevA.62.022703},
  url = {https://link.aps.org/doi/10.1103/PhysRevA.62.022703}
}

@article{PhysRevB.64.235118,
  title = {Density-functional perturbation theory with ultrasoft pseudopotentials},
  author = {Dal Corso, Andrea},
  journal = {Phys. Rev. B},
  volume = {64},
  issue = {23},
  pages = {235118},
  numpages = {17},
  year = {2001},
  month = {Nov},
  publisher = {American Physical Society},
  doi = {10.1103/PhysRevB.64.235118},
  url = {https://link.aps.org/doi/10.1103/PhysRevB.64.235118}
}

@article{PhysRevB.63.245101,
  title = {All-electron magnetic response with pseudopotentials: NMR chemical shifts},
  author = {Pickard, Chris J. and Mauri, Francesco},
  journal = {Phys. Rev. B},
  volume = {63},
  issue = {24},
  pages = {245101},
  numpages = {13},
  year = {2001},
  month = {5},
  publisher = {American Physical Society},
  doi = {10.1103/PhysRevB.63.245101},
  url = {https://link.aps.org/doi/10.1103/PhysRevB.63.245101}
}

@article{Todorov_2001,
doi = {10.1088/0953-8984/13/45/302},
url = {https://doi.org/10.1088/0953-8984/13/45/302},
year = {2001},
month = {oct},
publisher = {},
volume = {13},
number = {45},
pages = {10125},
author = {T N Todorov},
title = {Time-dependent tight binding},
journal = {Journal of Physics: Condensed Matter}
}

@article{REYES2002441,
title = {First principles dynamics of Li–He collisional excitation using atomic core potentials},
journal = {Chemical Physics Letters},
volume = {363},
number = {5},
pages = {441-446},
year = {2002},
issn = {0009-2614},
doi = {https://doi.org/10.1016/S0009-2614(02)01235-6},
url = {https://www.sciencedirect.com/science/article/pii/S0009261402012356},
author = {A. Reyes and D.A. Micha and K. Runge}
}

@article{PhysRevLett.91.196401,
  title = {Nonlocal Pseudopotentials and Magnetic Fields},
  author = {Pickard, Chris J. and Mauri, Francesco},
  journal = {Phys. Rev. Lett.},
  volume = {91},
  issue = {19},
  pages = {196401},
  numpages = {4},
  year = {2003},
  month = {11},
  publisher = {American Physical Society},
  doi = {10.1103/PhysRevLett.91.196401},
  url = {https://link.aps.org/doi/10.1103/PhysRevLett.91.196401}
}

@article{PhysRevB.69.235102,
  title = {Density-functional perturbational theory for dielectric tensors in the ultrasoft pseudopotential scheme},
  author = {Umari, P. and Gonze, Xavier and Pasquarello, Alfredo},
  journal = {Phys. Rev. B},
  volume = {69},
  issue = {23},
  pages = {235102},
  numpages = {11},
  year = {2004},
  month = {Jun},
  publisher = {American Physical Society},
  doi = {10.1103/PhysRevB.69.235102},
  url = {https://link.aps.org/doi/10.1103/PhysRevB.69.235102}
}

@article{PhysRevB.73.035408,
  title = {Time-dependent density functional theory with ultrasoft pseudopotentials: Real-time electron propagation across a molecular junction},
  author = {Qian, Xiaofeng and Li, Ju and Lin, Xi and Yip, Sidney},
  journal = {Phys. Rev. B},
  volume = {73},
  issue = {3},
  pages = {035408},
  numpages = {11},
  year = {2006},
  month = {Jan},
  publisher = {American Physical Society},
  doi = {10.1103/PhysRevB.73.035408},
  url = {https://link.aps.org/doi/10.1103/PhysRevB.73.035408}
}

@article{PhysRevB.76.024401,
  title = {Calculation of NMR chemical shifts for extended systems using ultrasoft pseudopotentials},
  author = {Yates, Jonathan R. and Pickard, Chris J. and Mauri, Francesco},
  journal = {Phys. Rev. B},
  volume = {76},
  issue = {2},
  pages = {024401},
  numpages = {11},
  year = {2007},
  month = {Jul},
  publisher = {American Physical Society},
  doi = {10.1103/PhysRevB.76.024401},
  url = {https://link.aps.org/doi/10.1103/PhysRevB.76.024401}
}

@misc{rostgaard2009projectoraugmentedwavemethod,
      title={The Projector Augmented-wave Method}, 
      author={Carsten Rostgaard},
      year={2009},
      eprint={0910.1921},
      archivePrefix={arXiv},
      primaryClass={cond-mat.mtrl-sci},
      url={https://arxiv.org/abs/0910.1921}, 
}

@article{PhysRevLett.105.123002,
  title = {Exact Factorization of the Time-Dependent Electron-Nuclear Wave Function},
  author = {Abedi, Ali and Maitra, Neepa T. and Gross, E. K. U.},
  journal = {Phys. Rev. Lett.},
  volume = {105},
  issue = {12},
  pages = {123002},
  numpages = {4},
  year = {2010},
  month = {Sep},
  publisher = {American Physical Society},
  doi = {10.1103/PhysRevLett.105.123002},
  url = {https://link.aps.org/doi/10.1103/PhysRevLett.105.123002}
}

@article{PhysRevA.82.060701,
  title = {Revised Born-Oppenheimer approach and a reprojection method for inelastic collisions},
  author = {Belyaev, Andrey K.},
  journal = {Phys. Rev. A},
  volume = {82},
  issue = {6},
  pages = {060701},
  numpages = {4},
  year = {2010},
  month = {Dec},
  publisher = {American Physical Society},
  doi = {10.1103/PhysRevA.82.060701},
  url = {https://link.aps.org/doi/10.1103/PhysRevA.82.060701}
}

@article{10.1063/1.3665031,
    author = {Fatehi, Shervin and Alguire, Ethan and Shao, Yihan and Subotnik, Joseph E.},
    title = {Analytic derivative couplings between configuration-interaction-singles states with built-in electron-translation factors for translational invariance},
    journal = {The Journal of Chemical Physics},
    volume = {135},
    number = {23},
    pages = {234105},
    year = {2011},
    month = {12},
    issn = {0021-9606},
    doi = {10.1063/1.3665031},
    url = {https://doi.org/10.1063/1.3665031},
}

@article{10.1063/1.4747540,
    author = {Patchkovskii, Serguei},
    title = {Electronic currents and Born-Oppenheimer molecular dynamics},
    journal = {The Journal of Chemical Physics},
    volume = {137},
    number = {8},
    pages = {084109},
    year = {2012},
    month = {08},
    issn = {0021-9606},
    doi = {10.1063/1.4747540},
    url = {https://doi.org/10.1063/1.4747540},
}

@article{10.1063/1.3700800,
    author = {Ojanperä, Ari and Havu, Ville and Lehtovaara, Lauri and Puska, Martti},
    title = {Nonadiabatic Ehrenfest molecular dynamics within the projector augmented-wave method},
    journal = {The Journal of Chemical Physics},
    volume = {136},
    number = {14},
    pages = {144103},
    year = {2012},
    month = {04},
    issn = {0021-9606},
    doi = {10.1063/1.3700800},
    url = {https://doi.org/10.1063/1.3700800}
}

@article{doi:10.1021/jz3006173,
author = {Fatehi, Shervin and Subotnik, Joseph E.},
title = {Derivative Couplings with Built-In Electron-Translation Factors: Application to Benzene},
journal = {The Journal of Physical Chemistry Letters},
volume = {3},
number = {15},
pages = {2039-2043},
year = {2012},
doi = {10.1021/jz3006173},
URL = {https://doi.org/10.1021/jz3006173},
eprint = {https://doi.org/10.1021/jz3006173}
}

@article{PhysRevA.85.012702,
  title = {Ab initio treatment of charge transfer in ion-molecule collisions based on one-electron wave functions},
  author = {Gab\'as, P. M. M. and Errea, L. F. and M\'endez, L. and Rabad\'an, I.},
  journal = {Phys. Rev. A},
  volume = {85},
  issue = {1},
  pages = {012702},
  numpages = {12},
  year = {2012},
  month = {Jan},
  publisher = {American Physical Society},
  doi = {10.1103/PhysRevA.85.012702},
  url = {https://link.aps.org/doi/10.1103/PhysRevA.85.012702}
}

@article{doi:10.1021/ct400641n,
author = {Akimov, Alexey V. and Prezhdo, Oleg V.},
title = {The PYXAID Program for Non-Adiabatic Molecular Dynamics in Condensed Matter Systems},
journal = {Journal of Chemical Theory and Computation},
volume = {9},
number = {11},
pages = {4959-4972},
year = {2013},
doi = {10.1021/ct400641n},
    note ={PMID: 26583414},

URL = { 
    
        https://doi.org/10.1021/ct400641n
    
    

},
eprint = { 
    
        https://doi.org/10.1021/ct400641n
    
    

}

}

@article{doi:10.1021/ct400700c,
author = {Scherrer, A. and Vuilleumier, R. and Sebastiani, D.},
title = {Nuclear Velocity Perturbation Theory of Vibrational Circular Dichroism},
journal = {Journal of Chemical Theory and Computation},
volume = {9},
number = {12},
pages = {5305-5312},
year = {2013},
doi = {10.1021/ct400700c},
    note ={PMID: 26592268},
URL = { 
        https://doi.org/10.1021/ct400700c
},
eprint = { 
        https://doi.org/10.1021/ct400700c
}
}

@article{PhysRevB.88.085117,
  title = {Optimized norm-conserving Vanderbilt pseudopotentials},
  author = {Hamann, D. R.},
  journal = {Phys. Rev. B},
  volume = {88},
  issue = {8},
  pages = {085117},
  numpages = {10},
  year = {2013},
  month = {8},
  publisher = {American Physical Society},
  doi = {10.1103/PhysRevB.88.085117},
  url = {https://link.aps.org/doi/10.1103/PhysRevB.88.085117}
}

@book{Grosso2013,
author = {Grosso , G. and Pastori Parravicini, G.},
title = {Solid State Physics},
year = {2013},
isbn = {978-0-471-49028-9},
publisher = {Academic Press},
address = {New York, USA},
edition = {2nd,},
}

@article{doi:10.1021/jp505767b,
author = {Alguire, Ethan C. and Ou, Qi and Subotnik, Joseph E.},
title = {Calculating Derivative Couplings between Time-Dependent Hartree–Fock Excited States with Pseudo-Wavefunctions},
journal = {The Journal of Physical Chemistry B},
volume = {119},
number = {24},
pages = {7140-7149},
year = {2015},
doi = {10.1021/jp505767b},
note ={PMID: 25148602},
URL = {https://doi.org/10.1021/jp505767b},
eprint = {https://doi.org/10.1021/jp505767b}
}

@article{PhysRevD.91.085009,
  title = {Third order equations of motion and the Ostrogradsky instability},
  author = {Motohashi, Hayato and Suyama, Teruaki},
  journal = {Phys. Rev. D},
  volume = {91},
  issue = {8},
  pages = {085009},
  numpages = {5},
  year = {2015},
  month = {Apr},
  publisher = {American Physical Society},
  doi = {10.1103/PhysRevD.91.085009},
  url = {https://link.aps.org/doi/10.1103/PhysRevD.91.085009}
}

@article{10.1063/1.4906941,
    author = {Ou, Qi and Bellchambers, Gregory D. and Furche, Filipp and Subotnik, Joseph E.},
    title = {First-order derivative couplings between excited states from adiabatic TDDFT response theory},
    journal = {The Journal of Chemical Physics},
    volume = {142},
    number = {6},
    pages = {064114},
    year = {2015},
    month = {02},
    issn = {0021-9606},
    doi = {10.1063/1.4906941},
    url = {https://doi.org/10.1063/1.4906941},
}

@article{10.1063/1.4928578,
    author = {Scherrer, Arne and Agostini, Federica and Sebastiani, Daniel and Gross, E. K. U. and Vuilleumier, Rodolphe},
    title = {Nuclear velocity perturbation theory for vibrational circular dichroism: An approach based on the exact factorization of the electron-nuclear wave function},
    journal = {The Journal of Chemical Physics},
    volume = {143},
    number = {7},
    pages = {074106},
    year = {2015},
    month = {08},
    issn = {0021-9606},
    doi = {10.1063/1.4928578},
    url = {https://doi.org/10.1063/1.4928578},
}

@article{VANSETTEN201839,
title = {The PseudoDojo: Training and grading a 85 element optimized norm-conserving pseudopotential table},
journal = {Computer Physics Communications},
volume = {226},
pages = {39-54},
year = {2018},
issn = {0010-4655},
doi = {https://doi.org/10.1016/j.cpc.2018.01.012},
url = {https://www.sciencedirect.com/science/article/pii/S0010465518300250},
author = {M.J. {van Setten} and M. Giantomassi and E. Bousquet and M.J. Verstraete and D.R. Hamann and X. Gonze and G.-M. Rignanese}
}

@article{doi:10.1021/acs.jpclett.0c03080,
author = {Chu, Weibin and Zheng, Qijing and Akimov, Alexey V. and Zhao, Jin and Saidi, Wissam A. and Prezhdo, Oleg V.},
title = {Accurate Computation of Nonadiabatic Coupling with Projector Augmented-Wave Pseudopotentials},
journal = {The Journal of Physical Chemistry Letters},
volume = {11},
number = {23},
pages = {10073-10080},
year = {2020},
doi = {10.1021/acs.jpclett.0c03080},
}

@article{doi:10.1021/acs.jpclett.0c03853,
author = {Chu, Weibin and Prezhdo, Oleg V.},
title = {Concentric Approximation for Fast and Accurate Numerical Evaluation of Nonadiabatic Coupling with Projector Augmented-Wave Pseudopotentials},
journal = {The Journal of Physical Chemistry Letters},
volume = {12},
number = {12},
pages = {3082-3089},
year = {2021},
doi = {10.1021/acs.jpclett.0c03853},
    note ={PMID: 33750138},

URL = { 
    
        https://doi.org/10.1021/acs.jpclett.0c03853
    
    

},
eprint = { 
    
        https://doi.org/10.1021/acs.jpclett.0c03853
    
    

}

}

@article{doi:10.1021/acs.jctc.2c00006,
author = {Ditler, Edward and Zimmermann, Tomáš and Kumar, Chandan and Luber, Sandra},
title = {Implementation of Nuclear Velocity Perturbation and Magnetic Field Perturbation Theory in CP2K and Their Application to Vibrational Circular Dichroism},
journal = {Journal of Chemical Theory and Computation},
volume = {18},
number = {4},
pages = {2448-2461},
year = {2022},
doi = {10.1021/acs.jctc.2c00006},
    note ={PMID: 35363490},
URL = {   
        https://doi.org/10.1021/acs.jctc.2c00006
},
eprint = {    
        https://doi.org/10.1021/acs.jctc.2c00006
}
}

@article{10.1063/5.0160965,
    author = {Athavale, Vishikh and Bian, Xuezhi and Tao, Zhen and Wu, Yanze and Qiu, Tian and Rawlinson, Jonathan and Littlejohn, Robert G. and Subotnik, Joseph E.},
    title = {Surface hopping, electron translation factors, electron rotation factors, momentum conservation, and size consistency},
    journal = {The Journal of Chemical Physics},
    volume = {159},
    number = {11},
    pages = {114120},
    year = {2023},
    month = {09},
    issn = {0021-9606},
    doi = {10.1063/5.0160965},
    url = {https://doi.org/10.1063/5.0160965},
}

@article{2023_Ditler_Mattiat_Luber,
author ="Ditler, Edward and Mattiat, Johann and Luber, Sandra",
title  ="The position operator problem in periodic calculations with an emphasis on theoretical spectroscopy",
journal  ="Phys. Chem. Chem. Phys.",
year  ="2023",
volume  ="25",
issue  ="21",
pages  ="14672-14685",
publisher  ="The Royal Society of Chemistry",
doi  ="10.1039/D2CP05991F",
url  ="http://dx.doi.org/10.1039/D2CP05991F"}

@article{10.1063/5.0230570,
    author = {Xu, Jianhang and Zhou, Ruiyi and Li, Tao E. and Hammes-Schiffer, Sharon and Kanai, Yosuke},
    title = {Lagrangian formulation of nuclear–electronic orbital Ehrenfest dynamics with real-time TDDFT for extended periodic systems},
    journal = {The Journal of Chemical Physics},
    volume = {161},
    number = {19},
    pages = {194109},
    year = {2024},
    month = {11},
    issn = {0021-9606},
    doi = {10.1063/5.0230570},
    url = {https://doi.org/10.1063/5.0230570},
}

@article{10.1063/5.0182685,
    author = {Mortensen, Jens J{\o}rgen and Larsen, Ask Hjorth and Kuisma, Mikael and Ivanov, Aleksei V. and Taghizadeh, Alireza and Peterson, Andrew and Haldar, Anubhab and Dohn, Asmus Ougaard and Sch{\"a}fer, Christian and J{\'o}nsson, Elvar {\"O}rn and Hermes, Eric D. and Nilsson, Fredrik Andreas and Kastlunger, Georg and Levi, Gianluca and J{\'o}nsson, Hannes and H{\"a}kkinen, Hannu and Fojt, Jakub and Kangsabanik, Jiban and S{\o}dequist, Joachim and Lehtom{\"a}ki, Jouko and Heske, Julian and Enkovaara, Jussi and Winther, Kirsten Tr{\o}strup and Dulak, Marcin and Melander, Marko M. and Ovesen, Martin and Louhivuori, Martti and Walter, Michael and Gjerding, Morten and Lopez-Acevedo, Olga and Erhart, Paul and Warmbier, Robert and W{\"u}rdemann, Rolf and Kaappa, Sami and Latini, Simone and Boland, Tara Maria and Bligaard, Thomas Knight and Skovhus, Thorbj{\o}rn and Susi, Toma and Maxson, Tristan and Rossi, Tuomas and Chen, Xi and Schmerwitz, Yorick Leonard A. and Schi{\o}tz, Jakob and Olsen, Thomas and Jacobsen, Karsten Wedel and Thygesen, Kristian Sommer},
    title = {GPAW: An open Python package for electronic structure calculations},
    journal = {The Journal of Chemical Physics},
    volume = {160},
    number = {9},
    pages = {092503},
    year = {2024},
    month = mar,
    issn = {0021-9606},
    doi = {10.1063/5.0182685},
    url = {https://doi.org/10.1063/5.0182685}
}

@article{doi:10.1021/acs.jctc.5c01082,
author = {Vogt, Jan-Robert and Schulz, Michael and Souza Mattos, Rafael and Barbatti, Mario and Persico, Maurizio and Granucci, Giovanni and Hutter, J{\"u}rg and Hehn, Anna},
title = {A Density Functional Theory and Semiempirical Framework for Trajectory Surface Hopping on Extended Systems},
journal = {Journal of Chemical Theory and Computation},
volume = {21},
number = {20},
pages = {10474-10488},
year = {2025},
doi = {10.1021/acs.jctc.5c01082},
note ={PMID: 41108029},
URL = {https://doi.org/10.1021/acs.jctc.5c01082},
eprint = {https://doi.org/10.1021/acs.jctc.5c01082}
}

@article{10.1063/5.0252559,
    author = {Zobač, Vladimír and Kuisma, Mikael and Larsen, Ask Hjorth and Rossi, Tuomas and Susi, Toma},
    title = {Ehrenfest dynamics with localized atomic-orbital basis sets within the projector augmented-wave method},
    journal = {The Journal of Chemical Physics},
    volume = {162},
    number = {12},
    pages = {124117},
    year = {2025},
    month = {03},
    issn = {0021-9606},
    doi = {10.1063/5.0252559},
    url = {https://doi.org/10.1063/5.0252559},
}

@article{doi:10.1021/acs.jpca.5c01344,
author = {Kumar, Ravi and Luber, Sandra},
title = {Calculation of Vibrational Circular Dichroism Spectra Employing Nuclear Velocity Perturbation or Magnetic Field Perturbation Theory Using an Atomic-Orbital-Based Linear Response Approach},
journal = {The Journal of Physical Chemistry A},
volume = {129},
number = {19},
pages = {4325-4336},
year = {2025},
doi = {10.1021/acs.jpca.5c01344},
    note ={PMID: 40310687},

URL = { 
    
        https://doi.org/10.1021/acs.jpca.5c01344
    
    

},
eprint = { 
    
        https://doi.org/10.1021/acs.jpca.5c01344
    
    

}

}

@article{PhysRevLett.136.196401,
  title = {Rototranslational Sum Rules for Nuclear Dynamics via Traveling Pseudopotentials},
  author = {Stengel, Massimiliano and Royo, Miquel and Artacho, Emilio},
  journal = {Phys. Rev. Lett.},
  volume = {136},
  issue = {19},
  pages = {196401},
  numpages = {7},
  year = {2026},
  month = {May},
  publisher = {American Physical Society},
  doi = {10.1103/sdwb-qxrp},
  url = {https://link.aps.org/doi/10.1103/sdwb-qxrp}
}

@article{z497-65ks,
  title = {Excess energy and countercurrents after a quantum kick},
  author = {Santerv\'as-Arranz, Nuria and Stengel, Massimiliano and Artacho, Emilio},
  journal = {Phys. Rev. Res.},
  volume = {7},
  issue = {3},
  pages = {033292},
  numpages = {10},
  year = {2025},
  month = {Sep},
  publisher = {American Physical Society},
  doi = {10.1103/z497-65ks},
  url = {https://link.aps.org/doi/10.1103/z497-65ks}
}

@misc{woodard2015theoremostrogradsky,
      title={The Theorem of Ostrogradsky}, 
      author={R. P. Woodard},
      year={2015},
      eprint={1506.02210},
      archivePrefix={arXiv},
      primaryClass={hep-th},
      url={https://arxiv.org/abs/1506.02210}, 
}

@article{doi:10.1021/acs.jctc.5c01960,
author = {Pu, Zhichen and Wu, Xiaojie and Wang, Yuanheng and Fan, Cheng and Yan, Wen and Zhou, Zehao and Gao, Yi Qin and Sun, Qiming},
title = {Analytical Excited-State Gradients and Derivative Couplings in TDDFT with Minimal Auxiliary Basis Set Approximation and GPU Acceleration},
journal = {Journal of Chemical Theory and Computation},
volume = {22},
number = {4},
pages = {1793-1810},
year = {2026},
doi = {10.1021/acs.jctc.5c01960},
    note ={PMID: 41615830},

URL = { 
    
        https://doi.org/10.1021/acs.jctc.5c01960
    
    

},
eprint = { 
    
        https://doi.org/10.1021/acs.jctc.5c01960
    
    

}

}

\end{document}